\documentclass[doublecol]{epl2}

\title{
Multipolar correlations and deformation effect on nuclear transition matrix
elements of double-$\beta $ decay}

\shorttitle{Multipolar correlations and deformation effect}

\author{
R. Chandra\inst{1,2} \and K. Chaturvedi\inst{3} \and P. K. Rath\inst{2} P. K.
Raina\inst{1} \and J. G. Hirsch\inst{4}}

\shortauthor{R. Chandra \etal}

\institute{
 \inst{1} Department of Physics and Meteorology, IIT Kharagpur,
          Kharagpur-721302, India\\
 \inst{2} Department of Physics, University of Lucknow, Lucknow-226007, India\\
 \inst{3} Department of Physics, Bundelkhand University, Jhansi-284128, India\\
 \inst{4} Instituto de Ciencias Nucleares, Universidad Nacional Aut\'{o}noma de M\'{e}xico,A.P. 70-543, M\'{e}xico 04510 D.F., M\'{e}xico
}


\abstract{
The two neutrino and neutrinoless double beta decay of $\ ^{94,96}$Zr, $%
^{98,100}$Mo, $^{104}$Ru, $^{110}$Pd, $^{128,130}$Te and $^{150}$Nd isotopes
for the $0^{+}\rightarrow 0^{+}$ transition is studied within the PHFB framework
along with an effective two-body interaction consisting of pairing, 
quadrupole-quadrupole and hexadecapole-hexadecapole correlations. It is 
found
that the effect of hexadecapolar correlations 
can be assimilated substantially as a renormalization of the
quadrupole-quadrupole 
interaction.
The effect of deformation on nuclear transition matrix elements 
is investigated by varying the strength of quadrupolar correlations in the parent and daughter nuclei
independently. The 
variation 
of the
nuclear transition matrix elements
as a function of the difference in deformation parameters of parent and
daughter nuclei reveals that in general, the former tend to be maximum for equal
deformation and they decrease as the difference in deformation parameters
increases,
exhibiting a very similar trend for the $\left( \beta^{-}\beta ^{-}\right)_{2\nu}$ 
and $\left( \beta^{-}\beta ^{-}\right)_{0\nu}$ transition matrix elements.
}

\begin{document}

\maketitle

The nuclear $\beta \beta $ decay is a rare second order semileptonic transition
between two even $Z$-even $N$ isobars $_{Z}^{A}X$ and $_{Z\pm 2}^{~~A}Y$
involving strangeness conserving weak charged currents. It is expected that
the nuclear $\beta \beta $ decay can proceed through sixteen experimentally
distinguishable modes, namely double-electron emission $\left( \beta
^{-}\beta ^{-}\right) $, double-positron emission $\left( \beta ^{+}\beta
^{+}\right) $, electron-positron conversion $\left( \varepsilon \beta
^{+}\right) $ and double-electron capture $\left( \varepsilon \varepsilon
\right) $ with the emission of two neutrinos, no neutrinos, single Majoron
and double Majorons. The $\beta^{+}\beta^{+}$, $\varepsilon\beta^{+}$ and 
$\varepsilon\varepsilon$ processes are energetically competing
and we shall refer to them as $e^{+}\beta \beta $ modes. The $\left( \beta
\beta \right) _{2\nu }$ decay conserves the lepton number and is an allowed
process in the standard model of electroweak unification (SM). On the other
hand, the $\left( \beta \beta \right) _{0\nu }$ decay violates the
conservation of lepton number and can occur in a number of gauge theoretical
models, namely GUTs -left-right symmetric models and E(6)-, R$_{p}$%
-conserving as well as violating SUSY models, in the scenarios of leptoquark
exchange, existence of heavy sterile neutrino, compositeness and Majoron
models. Hence, it is a convenient tool to test the physics beyond the SM.

The occurrence of $\left( \beta ^{-}\beta ^{-}\right) _{2\nu }$ decay was
proposed by Mayer in 1935 \cite{maye35} as the possibility to explain the
long observed half-lives of some even-even nuclei and it was finally
confirmed by Elliott $et$ $al.$ \cite{elli87} using a TPC detector.
Presently, the $\left( \beta ^{-}\beta ^{-}\right) _{2\nu }$ decay has been
already observed experimentally in 
ten nuclei, out of 35 possible candidates \cite{tret02}. 
For the
$\left( e^{+}\beta \beta \right) _{2\nu}$ modes, experimental limits on half-lives have been 
also given for 24 out of 34 possible isotopes \cite{tret02}. Klapdor-Kleingrothaus 
and his group have reported
the observation of ($\beta ^{-}\beta ^{-}$)$_{0\nu }$ decay of $^{76}$Ge
with $T_{1/2}^{0\nu }$ = 2.23$_{-0.31}^{+0.44}$ $\times $ 10$^{25}$ yr \cite
{klap01b, klap04}. The first results \cite{klap01b} were debated \cite
{feru02,aals02,zdes02,iann04}. 
These results could be confirmed by the forthcoming MAJORANA \cite{gehm06} and GERDA \cite{bett07}
experiments. 
It would be very exciting if this very first 
observation of ($\beta ^{-}\beta ^{-}$)$_{0\nu }$ decay by Klapdor-Kleingrothaus and his coworkers is verified 
by the latter two mentioned experiments.

It is possible to extract accurate nuclear transition matrix elements
(NTMEs) $M_{2\nu }$ from the observed half-lives of $\left( \beta ^{-}\beta
^{-}\right) _{2\nu }$ decay for the 0$^{+}\rightarrow 0^{+}$ transition and
a comparison between the theoretically calculated and experimentally
extracted NTMEs provides a cross-check on the reliability of different
nuclear models used for the study of nuclear $\beta \beta $ decay. Using the
average half-lives \cite{bara06} and the phase space factors \cite{boeh92},
the extracted NTMEs $M_{2\nu }$ vary from 0.0152$\pm $0.0008 (0.0238$\pm $%
0.0013) to 0.1222$\pm $0.0034 (0.1909$\pm $0.0053) for $g_{A}=1.25$ (1.00)
corresponding to $^{130}$Te and $^{100}$Mo respectively. In all cases, it is
observed that the NTMEs are sufficiently quenched. Vogel and Zirnbauer were
the first to provide an explanation of the observed suppression of $M_{2\nu
} $ in the QRPA 
employing a schematic surface-delta interaction
including the particle-particle channel, whose strength controls both
the concentration of $\beta ^{-}$ strength in the GTR and the
reduction of total $\beta ^{+}$ strength and its concentration at low
excitation energies \cite{voge86}. These conclusions were verified by
Civitarese $et$ $al.$ \cite{civi87} in the same QRPA model using realistic
interaction derived by G-matrix method. In the case of ($\beta ^{-}\beta
^{-} $)$_{0\nu }$ decay, the nuclear models usually predict half-lives
assuming certain value for the neutrino mass, or conversely extract various
parameters from the observed limits on half-lives. Over the past years, many
a nuclear models have been employed to study the nuclear $\beta \beta $
decay \cite{suho98,faes98}. The experimental as well as theoretical study of
nuclear $\beta \beta $ decay is quite wide in scope and has been excellently
reviewed over the past decades, which can be found in the recent review \cite
{avig08} and references there in.

The structure of a nucleus is mainly decided by the subtle interplay of
pairing and quadrupolar correlations present in the effective two-body
interaction. The former is responsible for the sphericity of the nucleus,
whereas the latter makes the nucleus deformed. In addition to the pairing
interaction, which plays an important role in all even $Z$-even $N$ $\beta
\beta $ emitters, it has been also shown that the deformation degrees of
freedom are crucial in the structure of $^{100}$Mo and $^{150}$Nd isotopes 
\cite{grif92}. The shell model is the best choice to calculate the NTMEs as
it attempts to solve the nuclear many-body problem as exactly as possible.
On the other hand, the QRPA and its extensions have emerged as the most
successful models in correlating single-$\beta $ GT strengths and half-lives
of ($\beta ^{-}\beta ^{-}$)$_{0\nu }$ decay. In spite of the spectacular
success of the QRPA in the study of $\beta \beta $ decay, there was a need to
include the deformation degrees of freedom in its formalism.

Over the past years, the deformed QRPA model has been developed for studying $\beta \beta $ decay
of spherical as well as deformed nuclei. The comparison of the experimental GT strength
distribution $B(GT)$ with the results of QRPA calculations was employed as a
novel method of deducing the deformation of $^{76}$Sr nucleus \cite{nach04}.
The effect of deformation on the $\left( \beta ^{-}\beta ^{-}\right) _{2\nu
} $ decay for the ground state transition $^{76}$Ge $\rightarrow $ $^{76}$Se
was studied in the framework of the deformed QRPA with separable GT residual
interaction \cite{pace04}. A deformed QRPA formalism to describe
simultaneously the energy distributions of the single-$\beta $ GT strength
and the $\left( \beta ^{-}\beta ^{-}\right) _{2\nu }$ decay matrix elements
of $^{48}$Ca, $^{76}$Ge, $^{82}$Se, $^{96}$Zr, $^{100}$Mo, $^{116}$Cd, $%
^{128,130}$Te, $^{136}$Xe and $^{150}$Nd isotopes using deformed Woods-Saxon
potentials and deformed Skyrme Hartree-Fock mean fields was developed \cite
{alva04}. Rodin and Faessler \cite{rodi08} have studied the $\beta \beta $
decay of $^{76}$Ge, $^{100}$Mo and $^{130}$Te isotopes and it has been
reported that the effect of continuum on the NTMEs of $\left( \beta
^{-}\beta ^{-}\right) _{2\nu }$ decay is negligible whereas the NTMEs of $%
\left( \beta ^{-}\beta ^{-}\right) _{0\nu }$ decay are regularly suppressed.

Recently, large scale calculations in the ``Interacting Shell Model'' (ISM)
have been performed to study the dependence of NTMEs for $\beta ^{-}\beta ^{-}$
decay of $^{48}$Ca, $^{76}$Ge, $^{82}$Se, $^{110}$Pd, $^{116}$Cd, $^{124}$%
Sn, $^{128,130}$Te and $^{136}$Xe nuclei on the valence space, effective
two-body interactions and deformation \cite{caur071}. In the ISM, the effect
of pairing and quadrupolar correlations on the NTMEs of $\left( \beta
^{-}\beta ^{-}\right) _{0\nu }$ decay has been also studied, in which it has
been noticed that the NTMEs are the largest in the presence of former only 
\cite{caur072,mene08}. It has been shown by Menendez \etal \cite{mene08}
that as the difference of deformation between parent and grand daughter grows,
the NTMEs for both $\left( \beta ^{-}\beta ^{-}\right) _{2\nu }$ and $\left(
\beta ^{-}\beta ^{-}\right) _{0\nu }$ decay decrease rapidly.

The PHFB model in conjunction with pairing plus quadrupole-quadrupole 
($PQQ$) interaction \cite{bara68} is a convenient choice to study the role of
quadrupolar correlations vis-a-vis deformation on the NTMEs for $\beta \beta 
$ decay. However, the PHFB model is unable to provide information about the
structure of the intermediate odd-odd nuclei in its present version and
hence, on the single $\beta $ decay rates and the distribution of GT
strength. In spite of this limitation, the PHFB model in conjunction with $%
PQQ$ interaction has been successfully applied to study the $\left( \beta
^{-}\beta ^{-}\right) _{2\nu }$ as well as $\left( e^{+}\beta \beta \right)
_{2\nu }$ decay modes for the $0^{+}\rightarrow 0^{+}$ transition where it
was possible to describe the lowest excited states of the parent and
daughter nuclei along with their electromagnetic transition strengths, as
well as to reproduce their measured $\beta \beta $ decay rates \cite
{chan05,rain06,sing07}.

An inverse correlation between the GT strength and quadrupole moment has
been noticed \cite{auer93}. In the PHFB model, the existence of an inverse
correlation between the quadrupole deformation and the size of NTMEs $%
M_{2\nu }$ has been also confirmed \cite{chan05,rain06,sing07}. In addition,
it has been observed that the NTMEs are usually large in the absence of
quadrupolar correlations. With the inclusion of the quadrupolar
correlations, the NTMEs are almost constant for small admixture of the $QQ$
interaction and suppressed substantially in realistic situation. In the mass
range $A=90-150$,
the suppression factor $D_{\alpha}$ for NTMEs 
evaluated without and with the quadrupole-quadrupole interaction ranges between 2 and 6, 
both for the $\left( \beta ^{-}\beta ^{-}\right)_{2\nu }$ and
$\left( \beta ^{-}\beta ^{-}\right)_{0\nu }$ decays.
%
For the double beta decay of $^{150}$%
Nd, it was 
also found  \cite{chat08}
that the NTMEs for $\left( \beta ^{-}\beta ^{-}\right)
_{0\nu }$ decay have a well defined maximum when the deformation of parent
and daughter nuclei are similar and they are suppressed for a difference in
the deformation.

The purpose of the present work is two fold. In the first part, we
investigate the effect of hexadecapolar correlations on the previously
calculated spectroscopic properties, NTMEs $M_{2\nu }$ for $\left( \beta
^{-}\beta ^{-}\right) _{2\nu }$ \cite{chan05,sing07} and $M_{0\nu }$ for $%
\left( \beta ^{-}\beta ^{-}\right) _{0\nu }$ decay \cite{chat08}. 
In the second part, we investigate the effect of having different
deformations in parent and daughter nuclei on NTMEs $M_{2\nu }$ and $M_{0\nu}$
of $\left(\beta ^{-}\beta ^{-}\right) _{2\nu }$ and 
$\left(\beta ^{-}\beta ^{-}\right) _{0\nu }$ decay respectively
for the $0^{+}\rightarrow 0^{+}$ transition.

The inverse half-life of the $\left( \beta ^{-}\beta ^{-}\right) _{2\nu }$\
decay for the$\ 0^{+}\to 0^{+}$\ transition is given by \cite
{haxt84,doi85,tomo91} 
\begin{equation}
\lbrack T_{1/2}^{2\nu }(0^{+}\to 0^{+})]^{-1}=G_{2\nu }|M_{2\nu }|^{2}
\end{equation}
The phase space factor $G_{2\nu }$ can be calculated with good accuracy \cite
{doi85,boeh92} and the nuclear model dependent NTME $M_{2\nu }$ is written
as 
\begin{eqnarray}
M_{2\nu } &=&\sum\limits_{N}\frac{\langle 0_{F}^{+}||\mathbf{\sigma }\tau
^{+}||1_{N}^{+}\rangle \langle 1_{N}^{+}||\mathbf{\sigma }\tau
^{+}||0_{I}^{+}\rangle }{E_{N}-(E_{I}+E_{F})/2}  \nonumber \\
&=&\sum\limits_{N}\frac{\langle 0_{F}^{+}||\mathbf{\sigma }\tau
^{+}||1_{N}^{+}\rangle \langle 1_{N}^{+}||\mathbf{\sigma }\tau
^{+}||0_{I}^{+}\rangle }{E_{0}+E_{N}-E_{I}}  \label{m2n}
\end{eqnarray}
with $\ $%
\begin{equation}
E_{0}=\frac{1}{2}\left( E_{I}-E_{F}\right) =\frac{1}{2}Q_{\beta \beta }+m_{e}
\end{equation}
The expressions to calculate $M_{2\nu }$ in the PHFB model 
are extensively discussed
in our earlier works \cite{chan05,sing07,chat08}. 
In what follows, we include only the required extensions in the formalism.
The HFB\ wave
functions are generated using an effective Hamiltonian with $PQQHH$ type of
effective two-body interaction. Explicitly, the Hamiltonian can be written
as 
\begin{equation}
H=H_{sp}+V(P)+\zeta _{qq}\left[ V(QQ)+V(HH)\right]
\end{equation}
where $H_{sp}$ denotes the single particle Hamiltonian. The $V(P)$, $V(QQ)$
and $V(HH)$ represent the pairing, quadrupole-quadrupole and
hexadecapole-hexadecapole part of the effective two-body interaction. The $\
\zeta _{qq}$ is an arbitrary parameter and the final results are obtained by
setting the$\ \zeta _{qq}=1$. The purpose of introducing $\zeta _{qq}$ is to
study the role of deformation by varying the strength of $QQHH$ interaction.

The pairing part of the effective two-body interaction $V(P)$ is given by 
\begin{equation}
V{(}P{)}=-\left( \frac{G}{4}\right) \sum\limits_{\alpha \beta
}(-1)^{j_{\alpha }+j_{\beta }-m_{\alpha }-m_{\beta }}a_{\alpha }^{\dagger
}a_{\bar{\alpha}}^{\dagger }a_{\bar{\beta}}a_{\beta }  \label{h}
\end{equation}
where $\alpha $ denotes the quantum numbers ($nljm$) and the state 
$\bar{\alpha}$ is same as $\alpha $ but with the sign of $m$ reversed. The $%
QQ$ part of the effective interaction $V(QQ)$\ is expressed as 
{\footnotesize 
\begin{equation}
V(QQ)=-\left( \frac{\chi _{2}}{2}\right) \sum\limits_{\alpha \beta \gamma
\delta }\sum\limits_{\mu }(-1)^{\mu }\langle \alpha |q_{2\mu }|\gamma
\rangle \langle \beta |q_{2-\mu }|\delta \rangle \ a_{\alpha }^{\dagger
}a_{\beta }^{\dagger }\ a_{\delta }\ a_{\gamma }  \label{i}
\end{equation}
} \noindent where 
\begin{equation}
q{_{2\mu }}=\left( \frac{16\pi }{5}\right) ^{1/2}r^{2}Y_{2\mu }(\theta ,\phi
)
\end{equation}
The $HH$ part of the effective interaction $V(HH)$ is given as 
{\footnotesize 
\begin{equation}
V(HH)=-\left( \frac{\chi _{4}}{2}\right) \sum\limits_{\alpha \beta \gamma
\delta }\sum\limits_{\nu }(-1)^{\nu }\langle \alpha |q_{4\nu }|\gamma
\rangle \langle \beta |q_{4-\nu }|\delta \rangle \ a_{\alpha }^{\dagger
}a_{\beta }^{\dagger }\ a_{\delta }\ a_{\gamma }
\end{equation}
} \noindent with 
\begin{equation}
q{_{4\nu }}=r^{4}Y_{4\nu }(\theta ,\phi )
\end{equation}
The relative magnitudes of the parameters of the $HH$ part of the two body
interaction are calculated from a relation suggested by Bohr and Mottelson 
\cite{bohr98}. According to them the approximate magnitude of these
constants for isospin $T=0$ is given by 
\begin{equation}
\chi _{\lambda }=\frac{4\pi }{2\lambda +1}\frac{m\omega _{0}^{2}}{%
A\left\langle r^{2\lambda -2}\right\rangle }\,\,\,\,\,\,\,\, {for }\mathrm{{%
\,}\lambda =1,2,3,4\cdot \cdot \cdot }
\end{equation}
and the parameters for the $T=1$ case are approximately half of their $T=0$
counterparts. 
The value of
$\chi _{4}$ for $T=1$ 
is taken
as exactly half of the $%
T=0$ case. Explicitely
\begin{eqnarray}
\chi _{4} &=&\left[ \left( \frac{16}{25}\right) \left( \frac{2}{3}\right)
^{2/3}\right] \chi _{2}A^{-2/3}b^{-4}  \nonumber \\
&=&0.4884\chi _{2}A^{-2/3}b^{-4}  \label{chi4}
\end{eqnarray}
with  $b=1.0032A^{1/6}$.

The model space,
single particle energies (SPE's), parameters of $PQQ$ type of effective
two-body interactions and the method to fix them are the same as in
references \cite{chan05,sing07,chat08}.
We fix $\chi _{2pn}$ as well as $\chi _{4pn}$ through the experimentally
available energy spectra for a given model space, SPE's, $G_{p}$, $G_{n},$ $%
\chi _{2pp}\,$and $\chi _{4pp}$. The strength of proton-neutron (\textit{pn}%
) component of the $QQ$ interaction $\chi _{2pn}$ is varied so as to obtain
the spectra of considered nuclei, namely $^{94,96}$Zr, $^{94,96,98,100}$Mo, $%
^{98,100,104}$Ru, $^{104,110}$Pd, $^{110}$Cd, $^{128,130}$Te, $^{128,130}$%
Xe, $^{150}$Nd and $^{150}$Sm in optimum agreement with the experimental
results. The theoretical spectra has been taken to be the optimum one if the
excitation energy of the $\ $2$^{+}$ state $E_{2^{+}}$ is reproduced as
closely as possible to the experimental value. All these input parameters
are kept fixed to calculate other spectroscopic properties.

In table~\ref{tab1}, we present the values of $\chi _{2pn}$ for the
$PQQ$ and $PQQHH$ interactions.
\begin{table}[ht]
\caption{Fitted values of $\chi _{2pn}$, calculated deformation parameters 
$\beta _{2}$ of parent(P) and daughter(D) nuclei participating in 
$\beta ^{-}\beta ^{-}$ decay using $PQQ$ and $PQQHH$ type of two-body 
interaction along with the average experimental values. 
$\dagger $ and $\ddagger $ denote $PQQ$ and $PQQHH$ interaction respectively.}
\label{tab1}
\begin{tabular}{llcll}
\hline\hline
&  & $\chi _{2pn}$ & \multicolumn{2}{c}{$\beta _{2}$} \\ 
&  &  & {\small Theory} & {\small Experiment\cite{rama01}} \\ \hline
$^{94}${\small Zr} & $\dagger $ & \multicolumn{1}{l}{\small 0.02519} & 
{\small 0.100} & {\small 0.090}$\pm ${\small 0.010} \\ 
& $\ddagger $ & \multicolumn{1}{l}{\small 0.02629} & {\small 0.110} &  \\ 
$^{94}${\small Mo} & $\dagger $ & \multicolumn{1}{l}{\small 0.02670} & 
{\small 0.161} & {\small 0.1509}$\pm ${\small 0.0015} \\ 
& $\ddagger $ & \multicolumn{1}{l}{\small 0.02572} & {\small 0.161} &  \\ 
$^{96}${\small Zr} & $\dagger $ & \multicolumn{1}{l}{\small 0.01717} & 
{\small 0.085} & {\small 0.080}$\pm ${\small 0.017} \\ 
& $\ddagger $ & \multicolumn{1}{l}{\small 0.01918} & {\small 0.087} &  \\ 
$^{96}${\small Mo} & $\dagger $ & \multicolumn{1}{l}{\small 0.02557} & 
{\small 0.191} & {\small 0.1720}$\pm ${\small 0.0016} \\ 
& $\ddagger $ & \multicolumn{1}{l}{\small 0.02472} & {\small 0.186} &  \\ 
$^{98}${\small Mo} & $\dagger $ & \multicolumn{1}{l}{\small 0.01955} & 
{\small 0.158} & {\small 0.1683}$\pm ${\small 0.0028} \\ 
& $\ddagger $ & \multicolumn{1}{l}{\small 0.01955} & {\small 0.162} &  \\ 
$^{98}${\small Ru} & $\dagger $ & \multicolumn{1}{l}{\small 0.02763} & 
{\small 0.205} & {\small 0.1947}$\pm ${\small 0.0030} \\ 
& $\ddagger $ & \multicolumn{1}{l}{\small 0.02649} & {\small 0.194} &  \\ 
$^{100}${\small Mo} & $\dagger $ & \multicolumn{1}{l}{\small 0.01906} & 
{\small 0.231} & {\small 0.2309}$\pm ${\small 0.0022} \\ 
& $\ddagger $ & \multicolumn{1}{l}{\small 0.01876} & {\small 0.226} &  \\ 
$^{100}${\small Ru} & $\dagger $ & \multicolumn{1}{l}{\small 0.01838} & 
{\small 0.214} & {\small 0.2148}$\pm ${\small 0.0011} \\ 
& $\ddagger $ & \multicolumn{1}{l}{\small 0.01831} & {\small 0.214} &  \\ 
$^{104}${\small Ru} & $\dagger $ & \multicolumn{1}{l}{\small 0.02110} & 
{\small 0.285} & {\small 0.2707}$\pm ${\small 0.0020} \\ 
& $\ddagger $ & \multicolumn{1}{l}{\small 0.02053} & {\small 0.282} &  \\ 
$^{104}${\small Pd} & $\dagger $ & \multicolumn{1}{l}{\small 0.01486} & 
{\small 0.216} & {\small 0.209}$\pm ${\small 0.007} \\ 
& $\ddagger $ & \multicolumn{1}{l}{\small 0.01507} & {\small 0.219} &  \\ 
$^{110}${\small Pd} & $\dagger $ & \multicolumn{1}{l}{\small 0.01417} & 
{\small 0.216} & {\small 0.257}$\pm ${\small 0.006} \\ 
& $\ddagger $ & \multicolumn{1}{l}{\small 0.01393} & {\small 0.214} &  \\ 
$^{110}${\small Cd} & $\dagger $ & \multicolumn{1}{l}{\small 0.01412} & 
{\small 0.196} & {\small 0.1770}$\pm ${\small 0.0039} \\ 
& $\ddagger $ & \multicolumn{1}{l}{\small 0.01414} & {\small 0.191} &  \\ 
$^{128}${\small Te} & $\dagger $ & \multicolumn{1}{l}{\small 0.02715} & 
{\small 0.136} & {\small 0.1363}$\pm ${\small 0.0011} \\ 
& $\ddagger $ & \multicolumn{1}{l}{\small 0.02692} & {\small 0.136} &  \\ 
$^{128}${\small Xe} & $\dagger $ & \multicolumn{1}{l}{\small 0.03600} & 
{\small 0.192} & {\small 0.1836}$\pm ${\small 0.0049} \\ 
& $\ddagger $ & \multicolumn{1}{l}{\small 0.02662} & {\small 0.181} &  \\ 
$^{130}${\small Te} & $\dagger $ & \multicolumn{1}{l}{\small 0.01801} & 
{\small 0.117} & {\small 0.1184}$\pm ${\small 0.0014} \\ 
& $\ddagger $ & \multicolumn{1}{l}{\small 0.01890} & {\small 0.120} &  \\ 
$^{130}${\small Xe} & $\dagger $ & \multicolumn{1}{l}{\small 0.02454} & 
{\small 0.166} & {\small 0.169}$\pm ${\small 0.007} \\ 
& $\ddagger $ & \multicolumn{1}{l}{\small 0.02281} & {\small 0.163} &  \\ 
$^{150}${\small Nd} & $\dagger $ & \multicolumn{1}{l}{\small 0.02160} & 
{\small 0.276} & {\small 0.2853}$\pm ${\small 0.0021} \\ 
& $\ddagger $ & \multicolumn{1}{l}{\small 0.02228} & {\small 0.279} &  \\ 
$^{150}${\small Sm} & $\dagger $ & \multicolumn{1}{l}{\small 0.01745} & 
{\small 0.238} & {\small 0.1931}$\pm ${\small 0.0021} \\ 
& $\ddagger $ & \multicolumn{1}{l}{\small 0.01730} & {\small 0.241} &  \\ 
\hline\hline
\end{tabular}
\end{table}
It is clear from the fitted values of $\chi _{2pn}$ that once
the procedure to fix $E_{2^{+}}$ is adopted, the change in the latter is
minimal even in the presence of $HH$ correlations. This behavior is further
reflected in the calculated yrast spectra, reduced $B(E2$:$0^{+}\to 2^{+})$
transition probabilities, static quadrupole moments $Q(2^{+})$ and
gyromagnetic factors $g(2^{+})$, which are usually in an overall agreement
with the experimental data \cite{chan05,sing07}. The maximum change in all
the calculated spectroscopic properties is about 10\% except for the $B(E2$:$%
0^{+}\to 2^{+})$ and $Q(2^{+})$ of $^{94}$Zr, which change by 20\% and 23\%
respectively. Further, we tabulate the theoretically calculated and
experimental deformation parameter $\beta _{2}$ \cite{rama01} of parent and
daughter nuclei involved in $\beta ^{-}\beta ^{-}$ decay in the same 
table~\ref{tab1} at $\zeta _{qq}=1$.
The calculated NTMEs $M_{2\nu }$ for the $0^{+}\rightarrow
0^{+}$ transition at $\zeta _{qq}=1$ are given in table~\ref{tab2}.
\begin{table}[ht]
\caption{
Calculated NTMEs $M_{2\nu }$ and $M_{0\nu }$ for $\left( \beta
^{-}\beta ^{-}\right) _{2\nu }$ and $\left( \beta ^{-}\beta ^{-}\right)
_{0\nu }$ decay respectively using $PQQ$ \cite{chan05,sing07,chat08} and $PQQHH$ interaction 
along with the extracted $M_{2\nu }$ from average/recommended 
experimental half-lives for the $0^{+}\rightarrow 0^{+}$
transition \cite{bara06}, except for $^{94}$Zr \cite{arno99}, and $^{110}$Pd 
\cite{wint52}. The numbers corresponding to (a) and (b) are calculated for $%
g_{A}=1.25$ and 1.0 respectively. $\dagger $ and $\ddagger $ denote $PQQ$
and $PQQHH$ interaction respectively.
}
\label{tab2}
\begin{tabular}{llcccc}
\hline\hline
&  & \multicolumn{3}{c}{$M_{2\nu }$} & $\left| M_{0\nu }\right| $ \\ 
&  & \multicolumn{1}{l}{\small Theory} & \multicolumn{2}{c}{\small Experiment%
} & {\small Theory} \\ \hline
&  &  & {\small (a)} & {\small (b)} &  \\ \hline
$^{94}${\small Zr} & $\dagger $ & \multicolumn{1}{l}{\small 0.076} & $<$%
{\small 62.8} & $<${\small 98.1} & {\small 2.034} \\ 
& $\ddagger $ & \multicolumn{1}{l}{\small 0.063} &  &  & {\small 1.866} \\ 
$^{96}${\small Zr} & $\dagger $ & \multicolumn{1}{l}{\small 0.058} & {\small %
0.051} & {\small 0.080} & {\small 1.453} \\ 
& $\ddagger $ & \multicolumn{1}{l}{\small 0.059} &  &  & {\small 1.425} \\ 
$^{98}${\small Mo} & $\dagger $ & \multicolumn{1}{l}{\small 0.130} &  &  & 
{\small 3.366} \\ 
& $\ddagger $ & \multicolumn{1}{l}{\small 0.127} &  &  & {\small 3.099} \\ 
$^{100}${\small Mo} & $\dagger $ & \multicolumn{1}{l}{\small 0.104} & 
{\small 0.122} & {\small 0.191} & {\small 3.252} \\ 
& $\ddagger $ & \multicolumn{1}{l}{\small 0.104} &  &  & {\small 3.080} \\ 
$^{104}${\small Ru} & $\dagger $ & \multicolumn{1}{l}{\small 0.068} &  &  & 
{\small 2.347} \\ 
& $\ddagger $ & \multicolumn{1}{l}{\small 0.068} &  &  & {\small 2.140} \\ 
$^{110}${\small Pd} & $\dagger $ & \multicolumn{1}{l}{\small 0.133} & $<$%
{\small 6.47} & $<${\small 10.11} & {\small 3.849} \\ 
& $\ddagger $ & \multicolumn{1}{l}{\small 0.120} &  &  & {\small 3.198} \\ 
$^{128}${\small Te} & $\dagger $ & \multicolumn{1}{l}{\small 0.033} & 
{\small 0.022} & {\small 0.034} & {\small 1.625} \\ 
& $\ddagger $ & \multicolumn{1}{l}{\small 0.033} &  &  & {\small 1.750} \\ 
$^{130}${\small Te} & $\dagger $ & \multicolumn{1}{l}{\small 0.042} & 
{\small 0.015} & {\small 0.024} & {\small 2.216} \\ 
& $\ddagger $ & \multicolumn{1}{l}{\small 0.031} &  &  & {\small 1.814} \\ 
$^{150}${\small Nd} & $\dagger $ & \multicolumn{1}{l}{\small 0.033} & 
{\small 0.033} & {\small 0.051} & {\small 1.616} \\ 
& $\ddagger $ & \multicolumn{1}{l}{\small 0.026} &  &  & {\small 1.210} \\ 
\hline\hline
\end{tabular}
\end{table}
The change in results for $\beta _{2}$
are below 10$\%$. The NTMEs $M_{2\nu } $ also change up to 10\% except $%
^{94} $Zr, $^{130}$Te and $^{150}$Nd, for which the changes are 17$\%$, 26$%
\% $ and 21$\%$ respectively. The ratio for deformation effect $D_{2\nu }$
are 2.29(2.23), 3.70(3.38), 1.86(1.75), 2.33(2.09), 5.47(5.19), 3.14(3.30),
4.26(3.97), 2.89(3.31) and 5.94(6.46) for $^{94,96}$Zr, $^{98,100}$Mo, $%
^{104}$Ru, $^{110}$Pd, $^{128,130}$Te and $^{150}$Nd nuclei with $PQQ$($%
PQQHH $) interaction respectively and their variation due to the $PQQHH$
type of interaction is up to 15\%. 

\begin{figure}[h]
\begin{tabular}{cc}
\resizebox{0.215\textwidth}{!}{\includegraphics{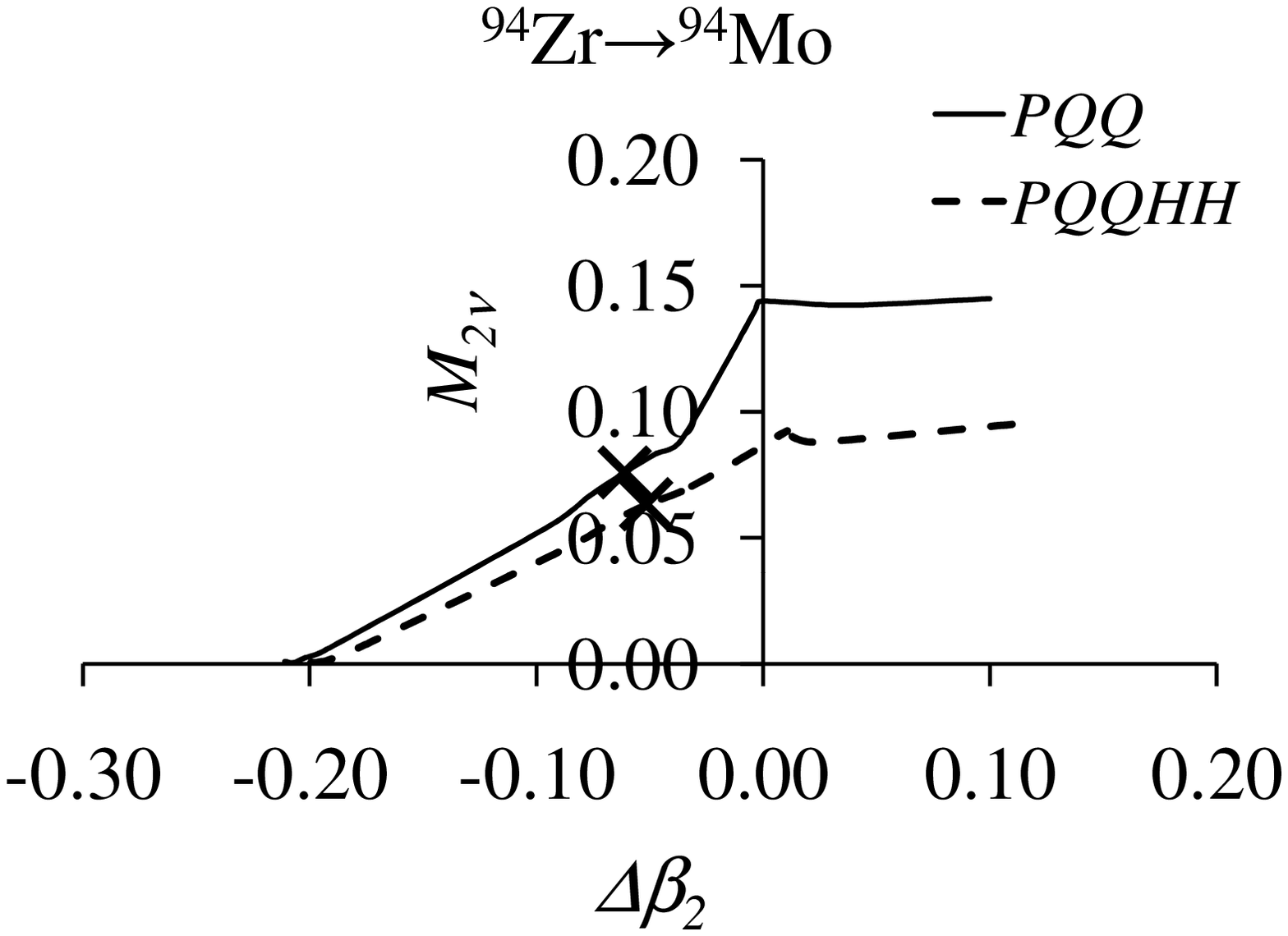}} &
\resizebox{0.215\textwidth}{!}{\includegraphics{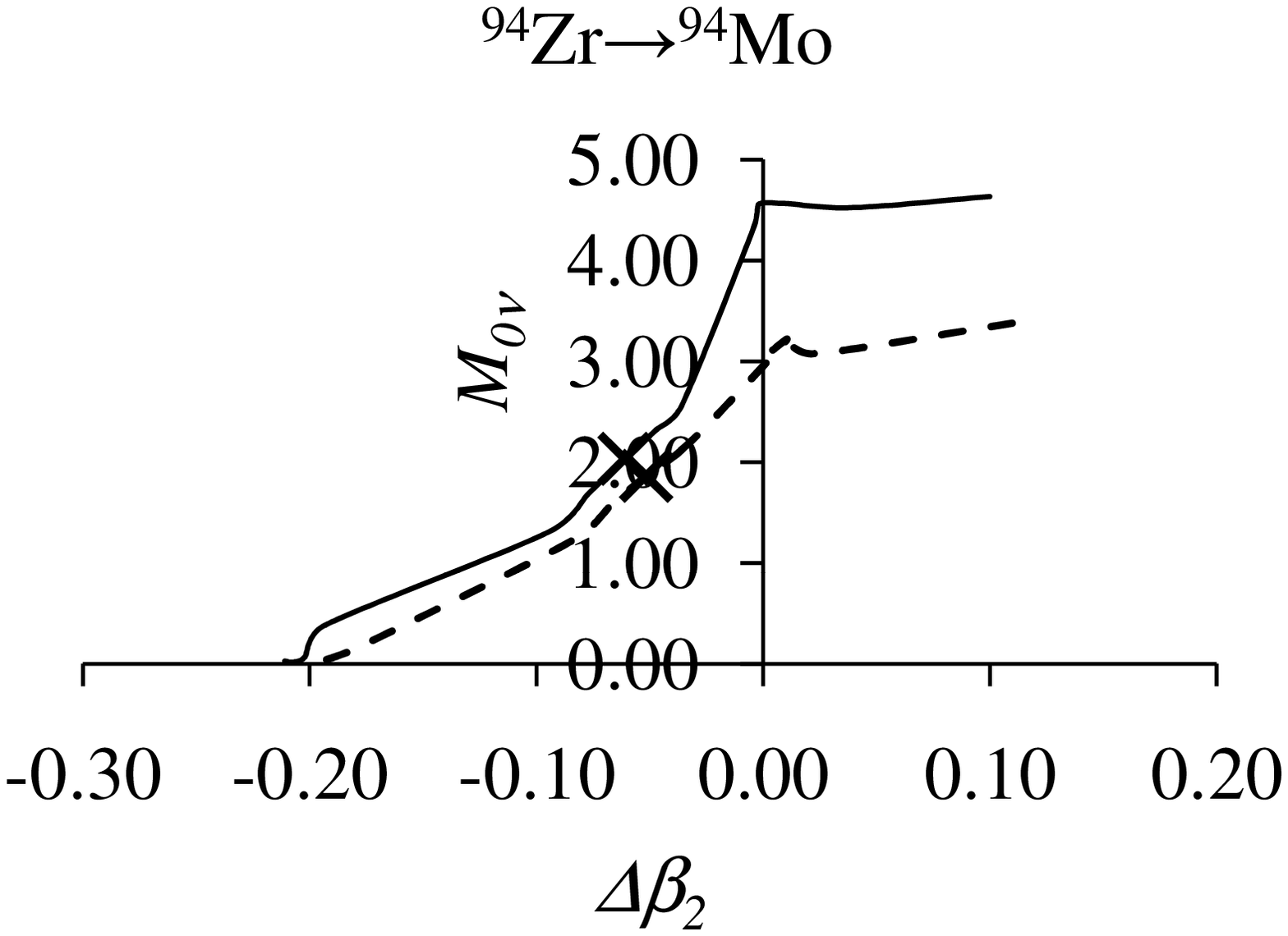}} \\
\resizebox{0.215\textwidth}{!}{\includegraphics{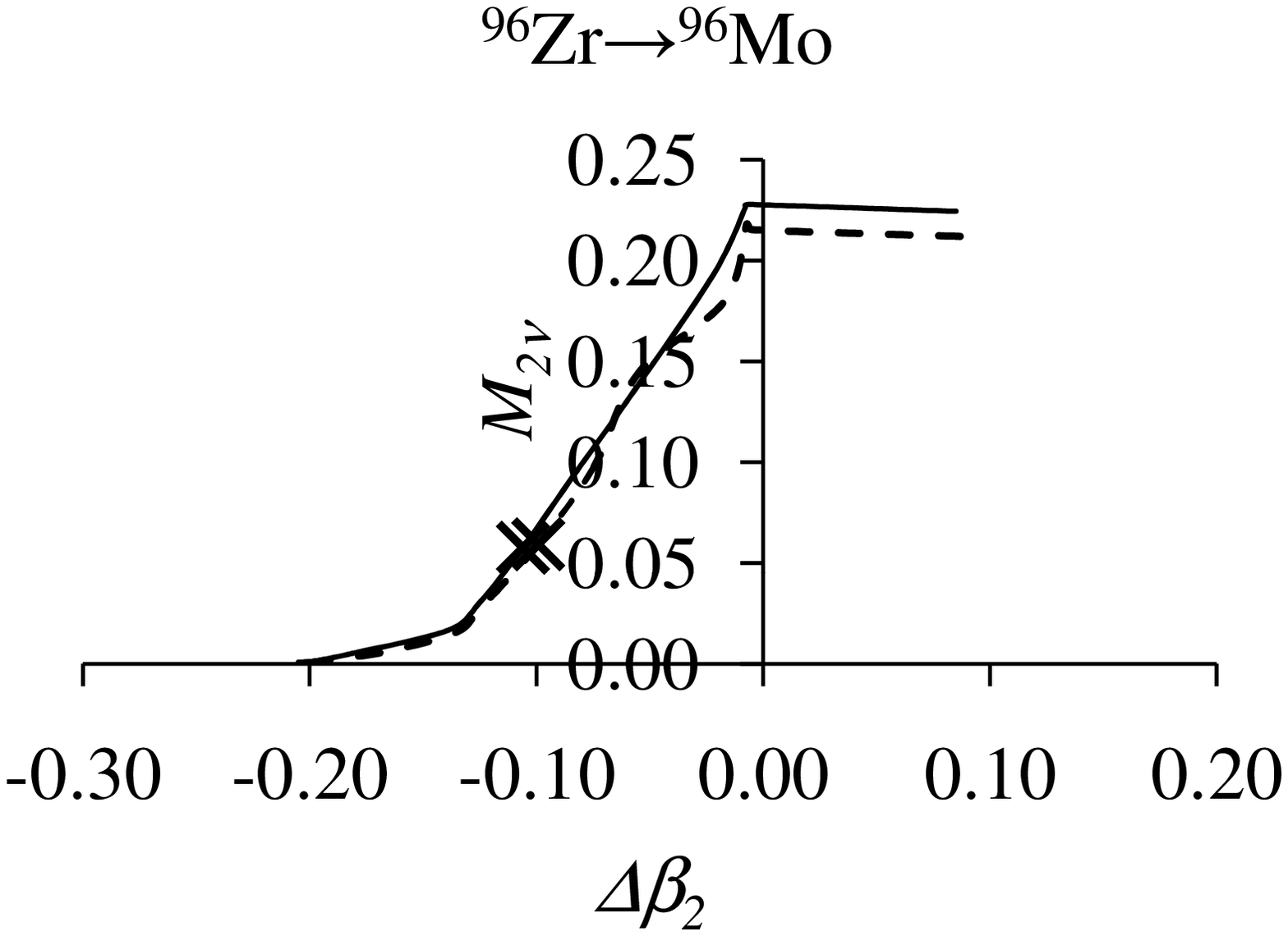}} &
\resizebox{0.215\textwidth}{!}{\includegraphics{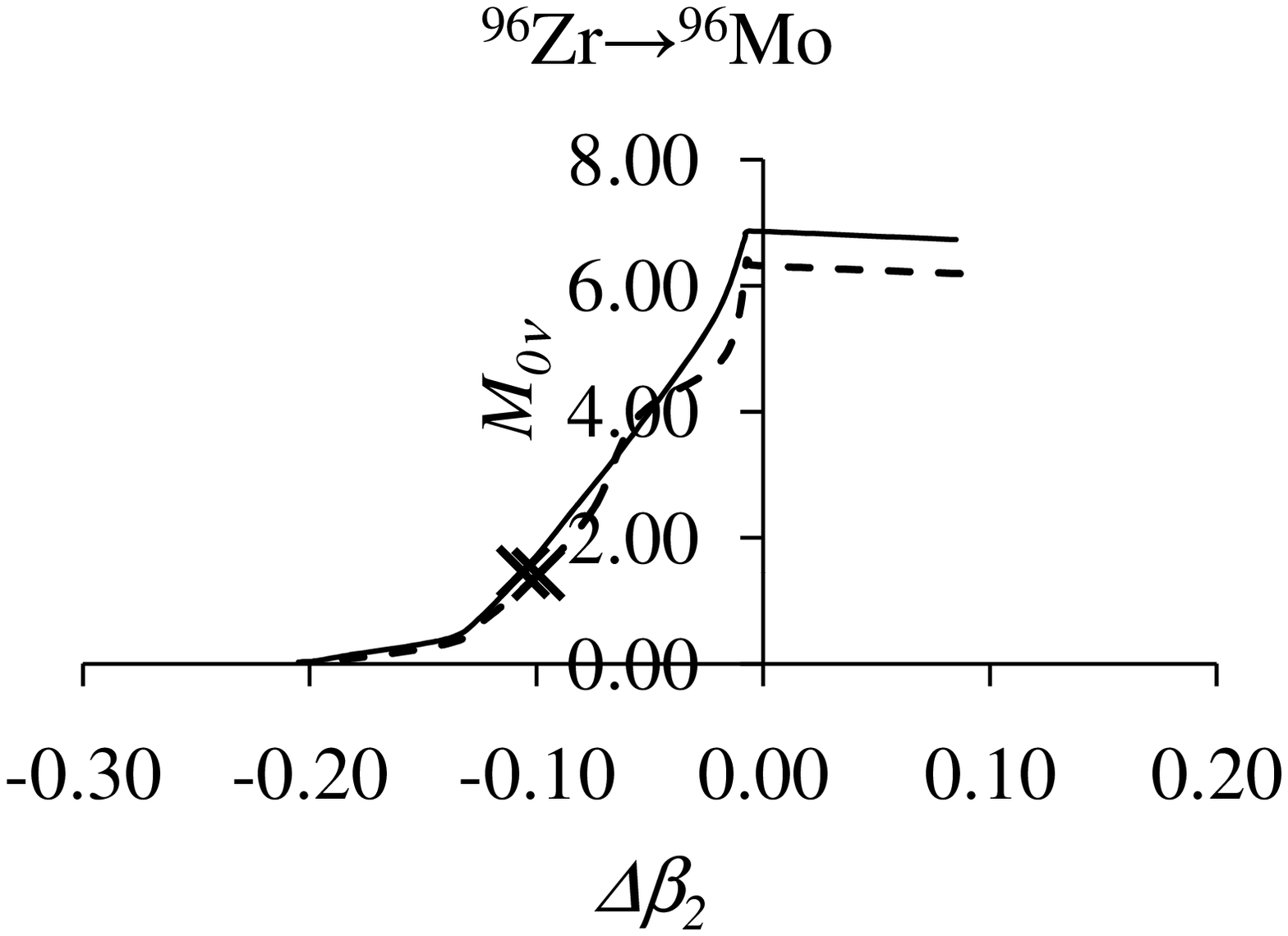}} \\
\resizebox{0.215\textwidth}{!}{\includegraphics{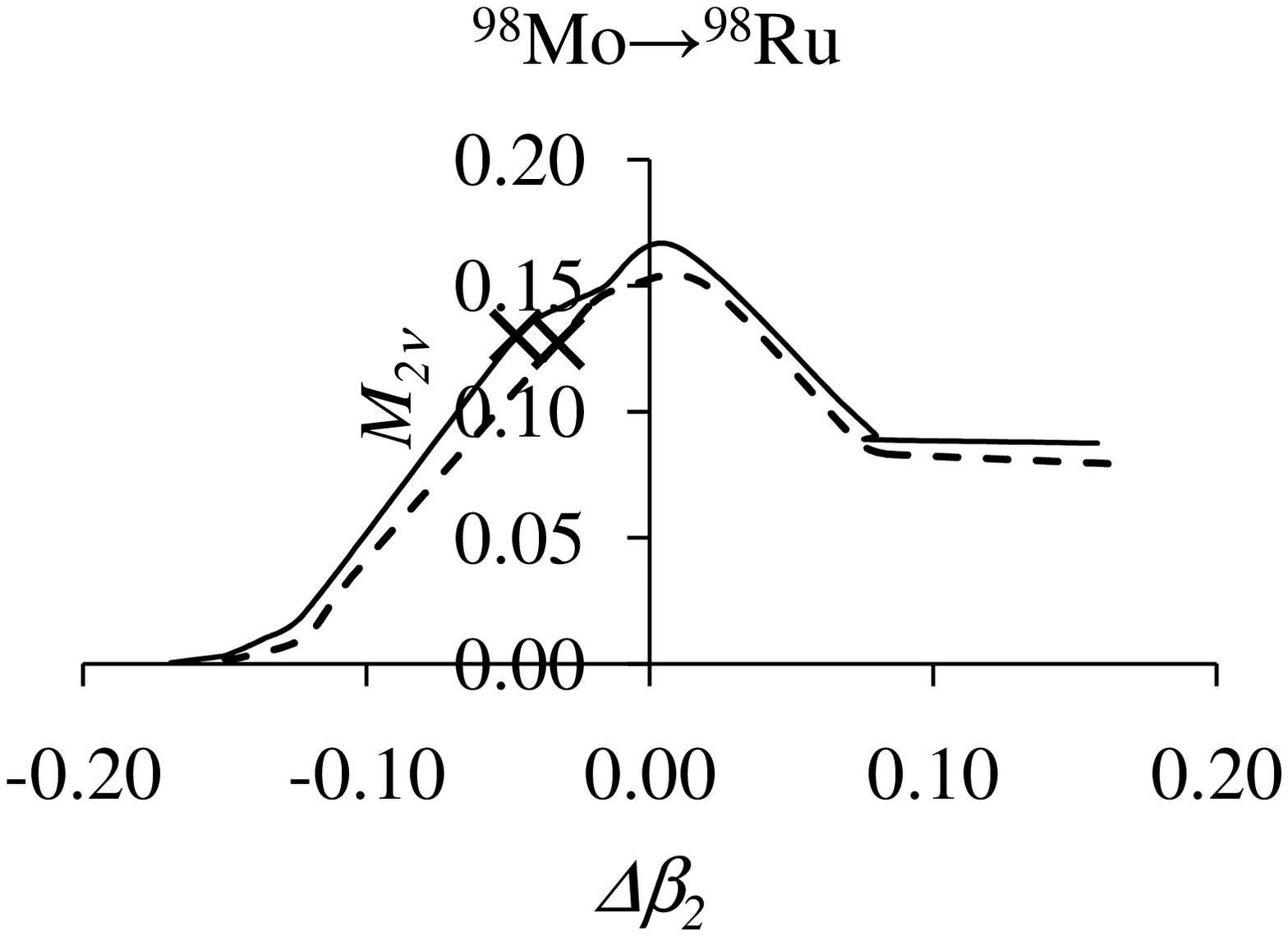}} &
\resizebox{0.215\textwidth}{!}{\includegraphics{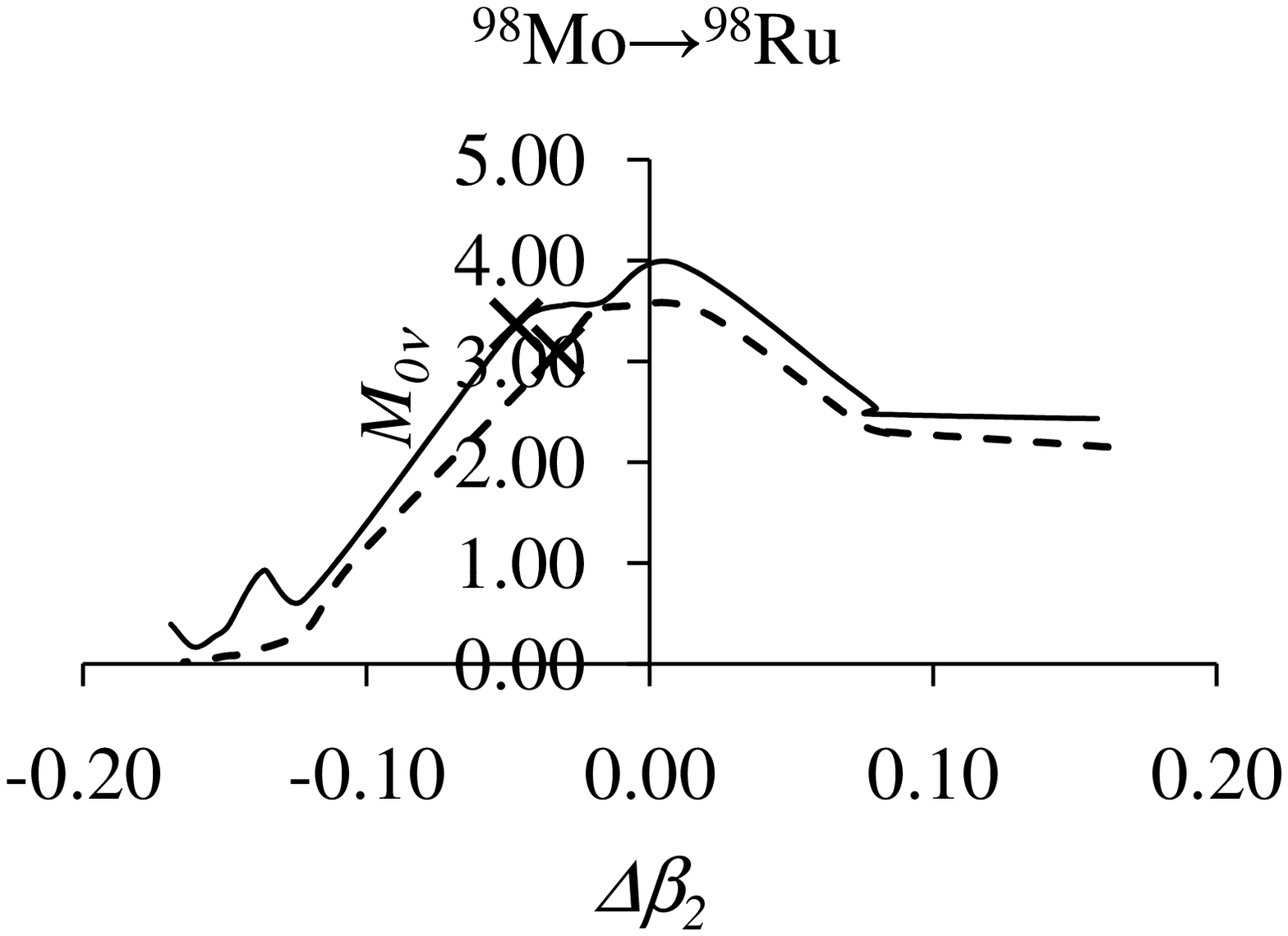}} \\
\resizebox{0.215\textwidth}{!}{\includegraphics{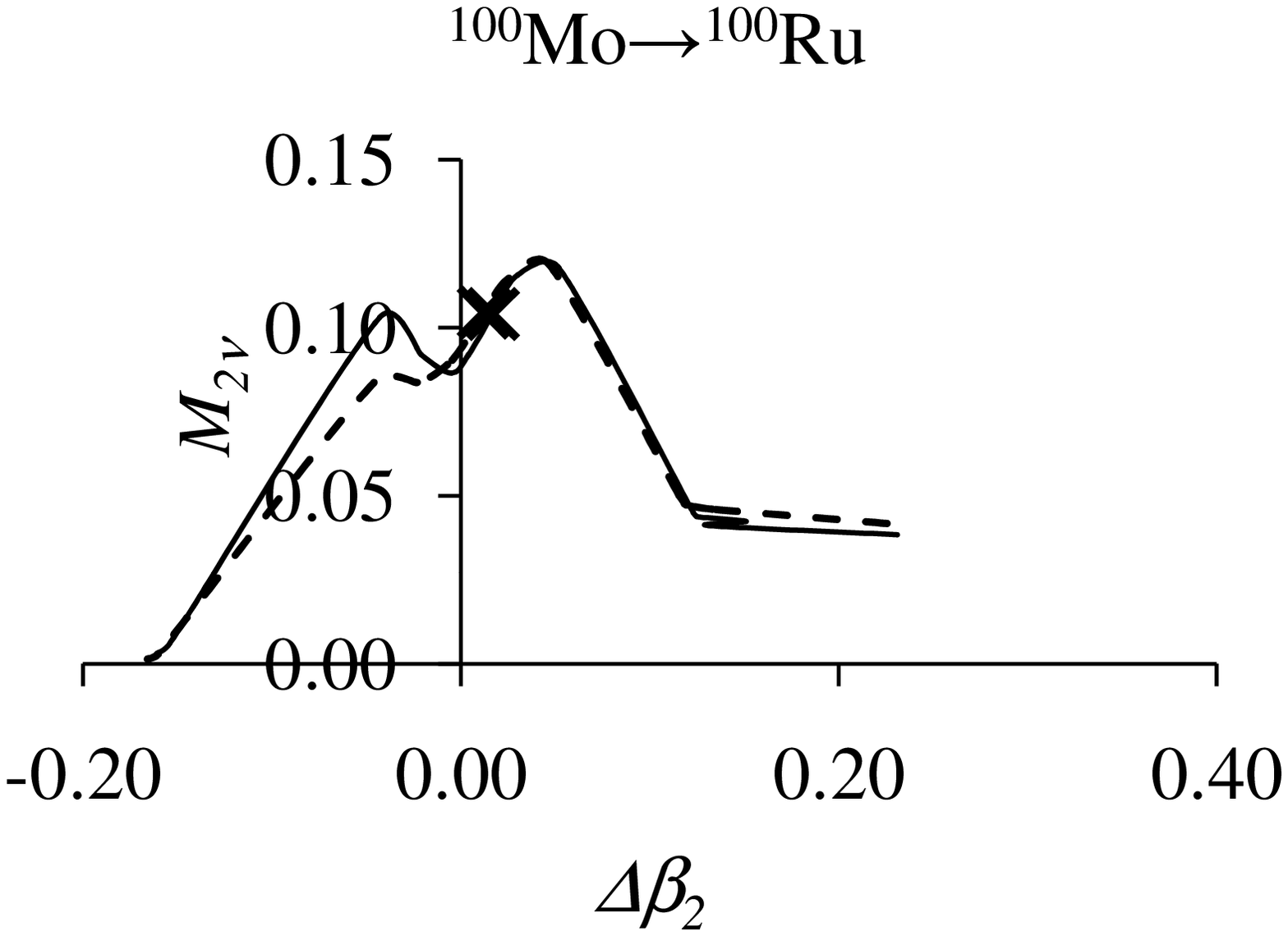}} &
\resizebox{0.215\textwidth}{!}{\includegraphics{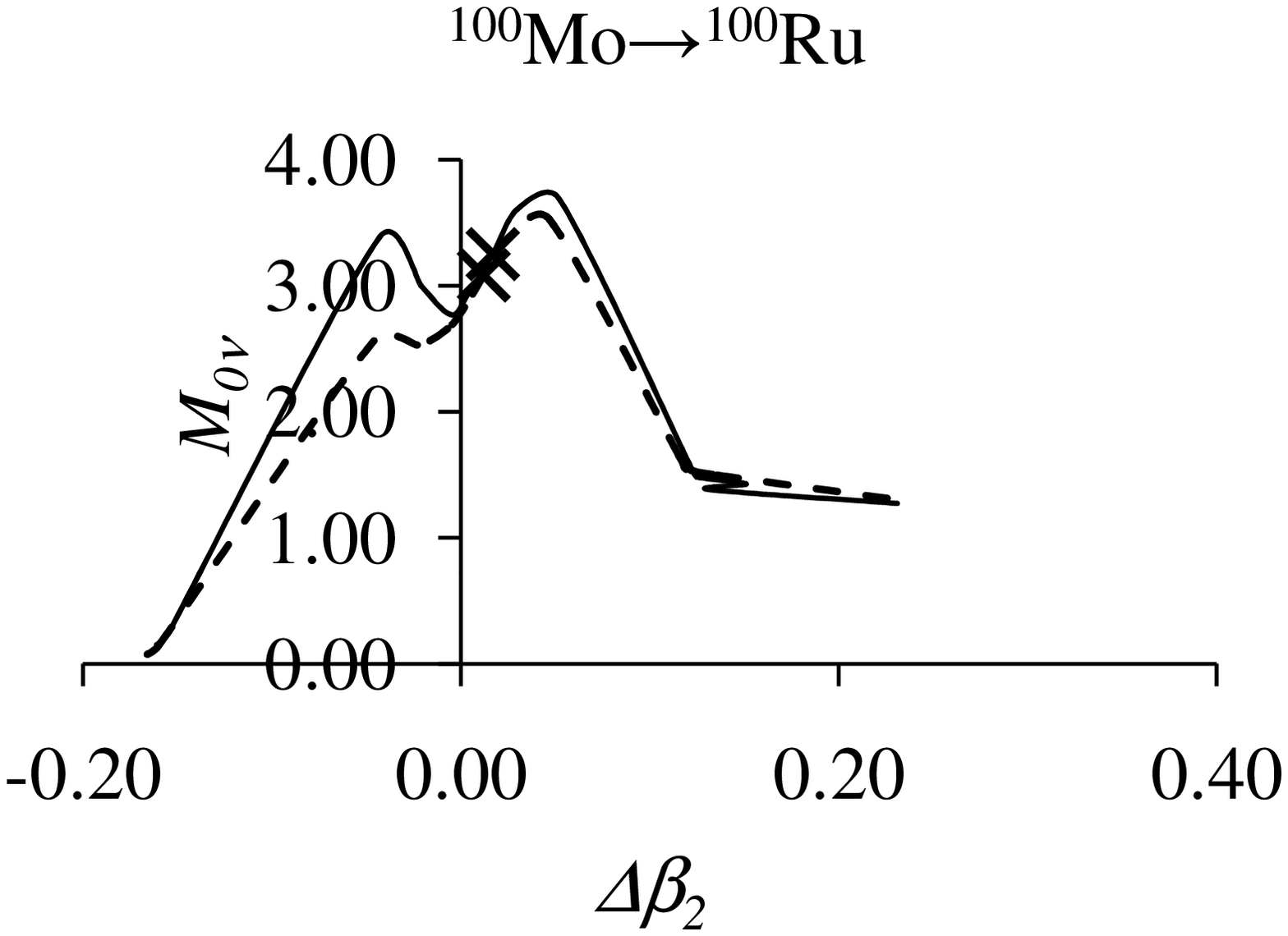}} \\
\resizebox{0.215\textwidth}{!}{\includegraphics{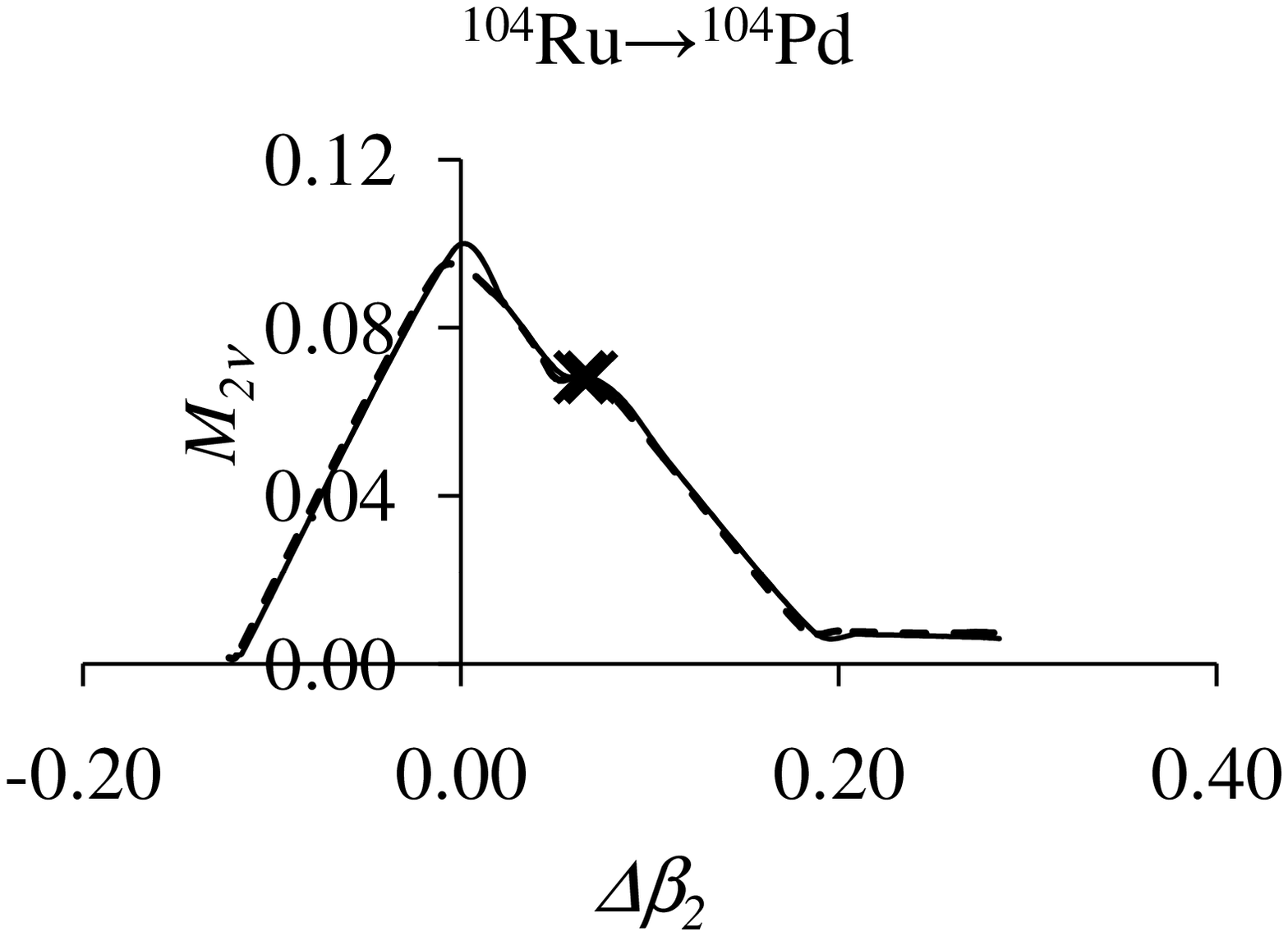}} &
\resizebox{0.215\textwidth}{!}{\includegraphics{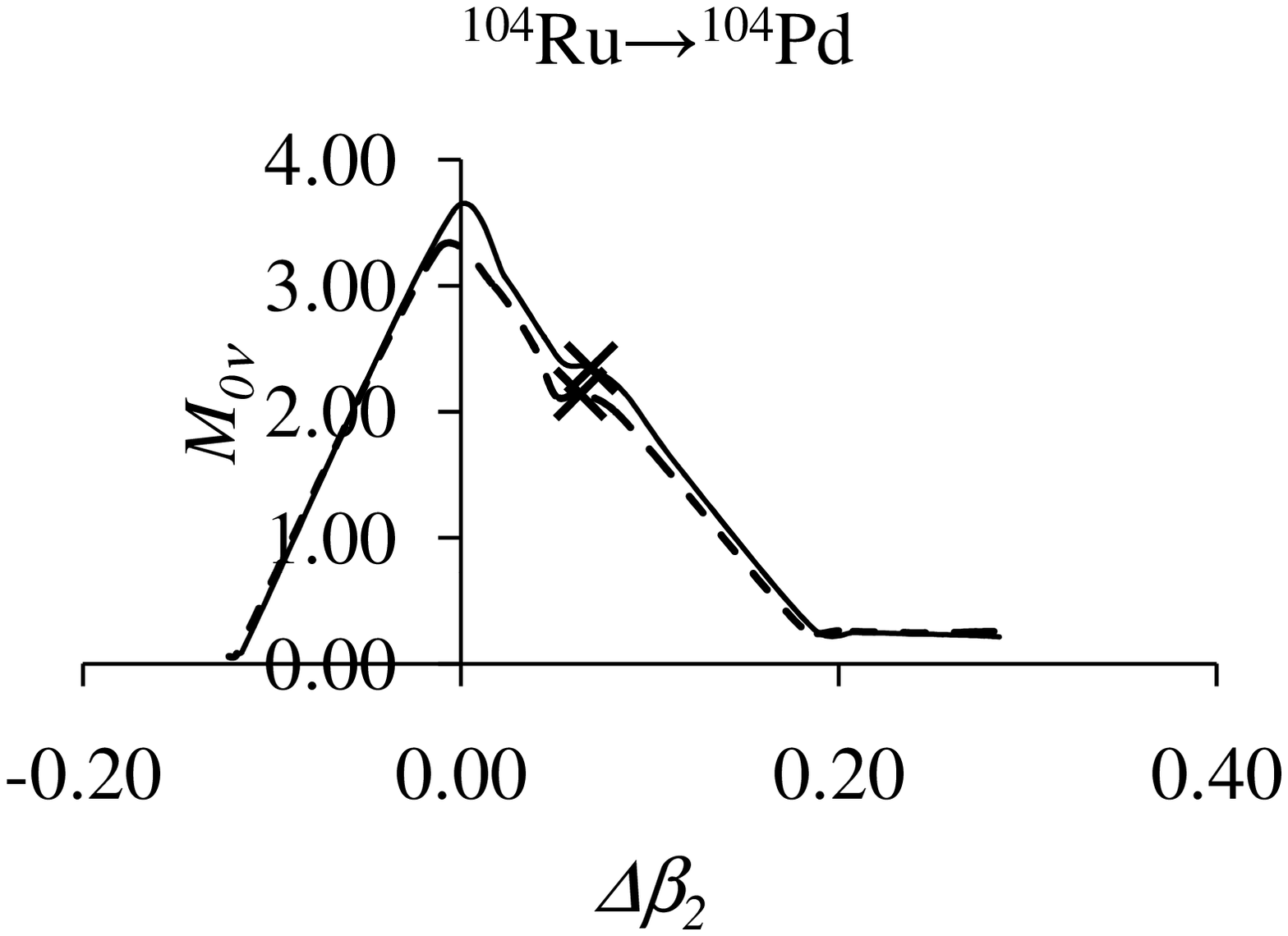}} \\
\resizebox{0.215\textwidth}{!}{\includegraphics{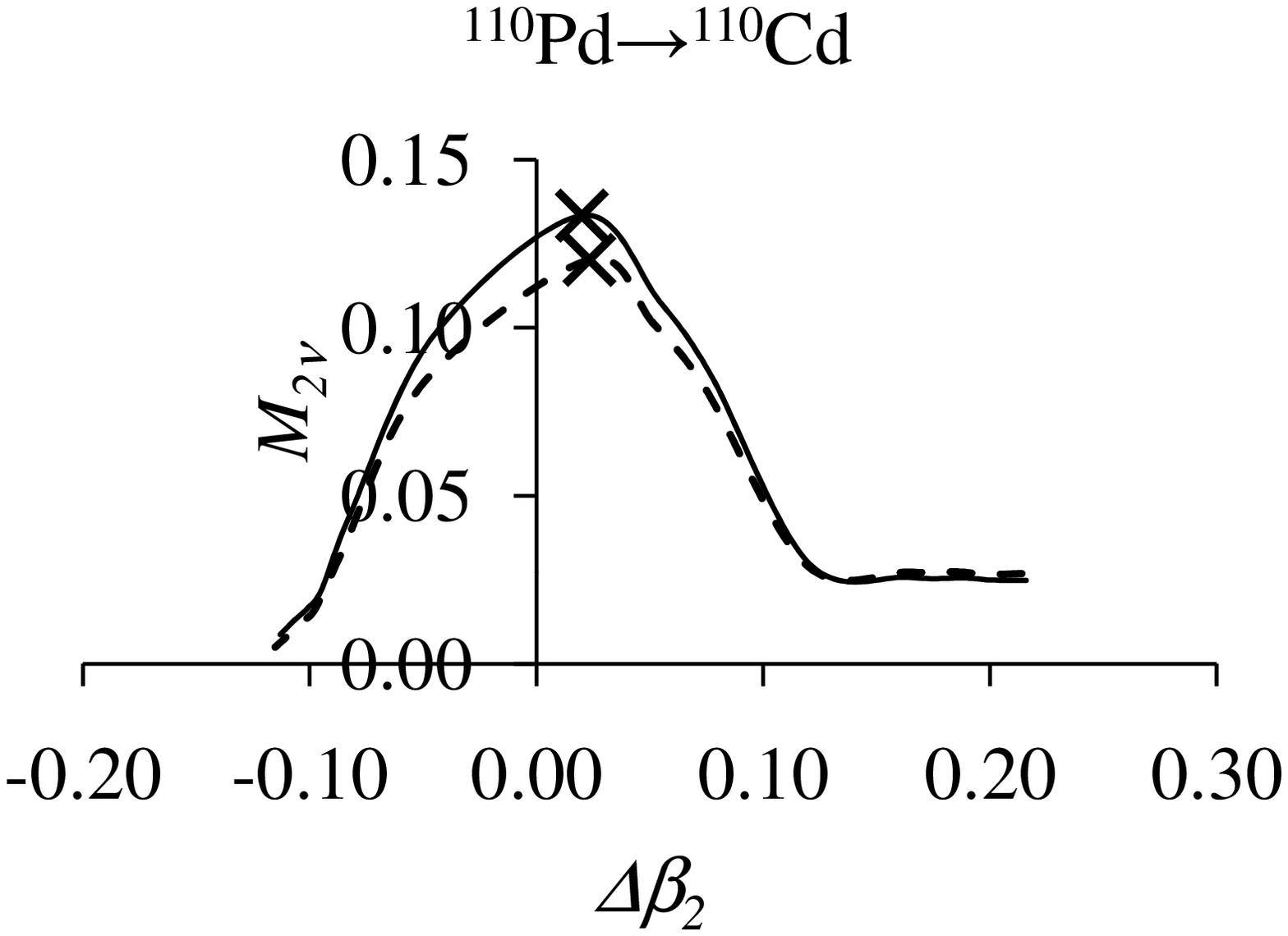}} &
\resizebox{0.215\textwidth}{!}{\includegraphics{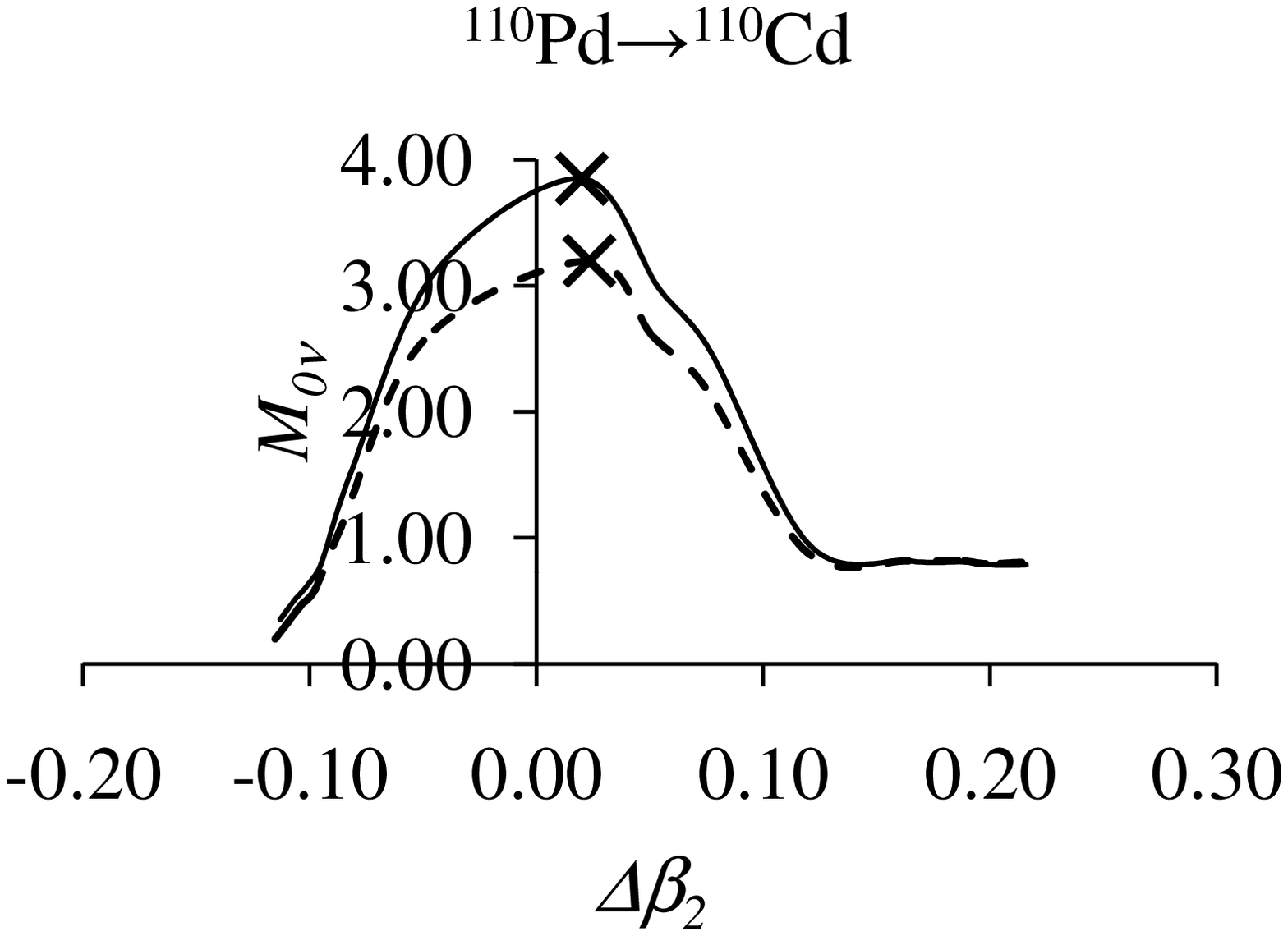}}
\end{tabular}
\caption{NTMEs of $^{94,96}$Zr, $^{98,100}$Mo, $^{104}$Ru, $^{110}$Pd 
isotopes for $\left(\beta^{-}\beta^{-}\right)_{2\nu}$ (left hand side) and
$\left(\beta^{-}\beta^{-}\right)_{0\nu}$ (right hand side) decay as a function of the
difference in the deformation parameter $\Delta\beta_2$. ``$\times$"
denotes the value of NTME for calculated $\Delta\beta_2$ from 
table~\ref{tab1}.}
\label{fig1}
\end{figure}

\begin{figure}[h]
\begin{tabular}{cc}
\resizebox{0.215\textwidth}{!}{\includegraphics{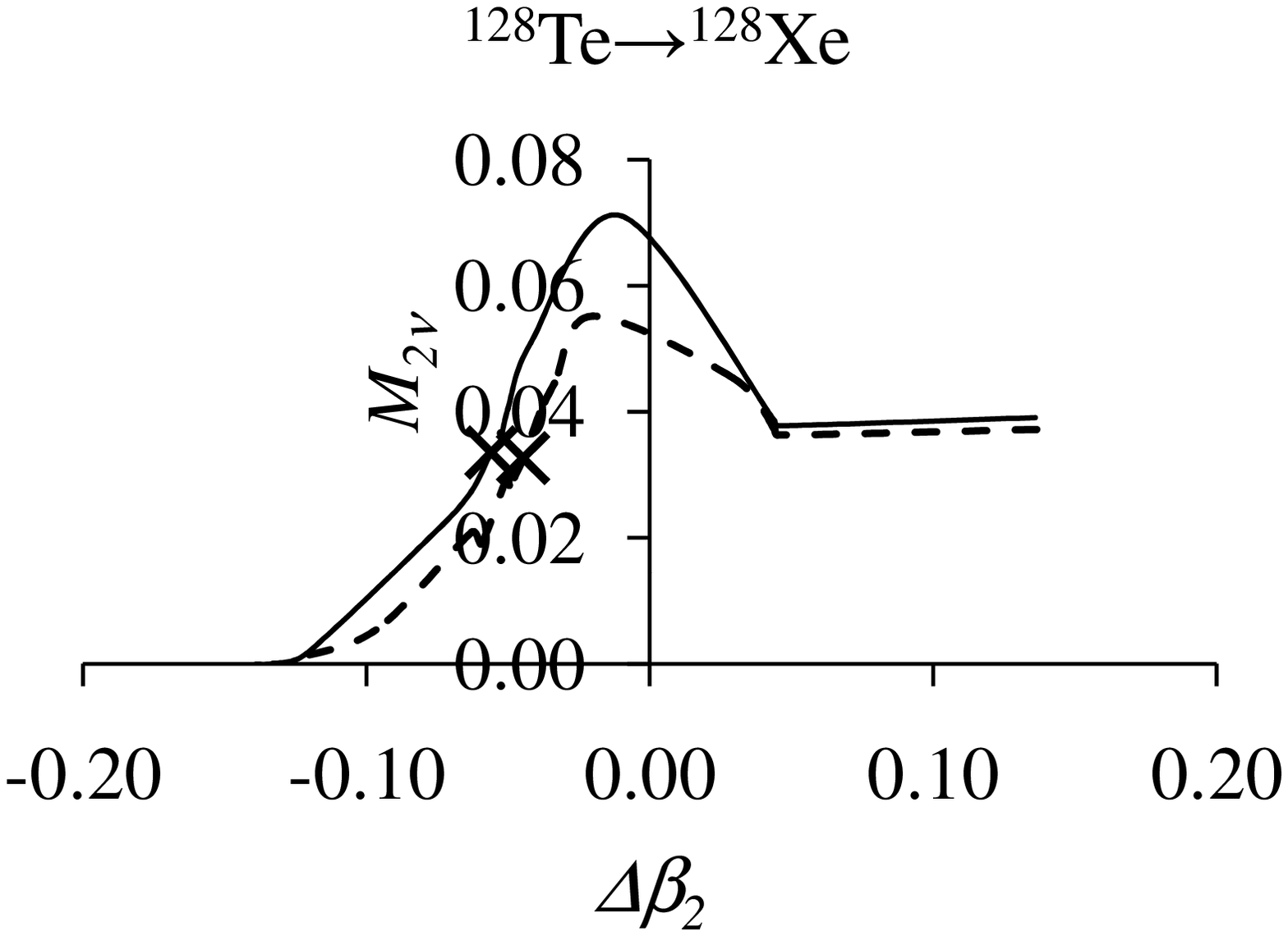}} &
\resizebox{0.215\textwidth}{!}{\includegraphics{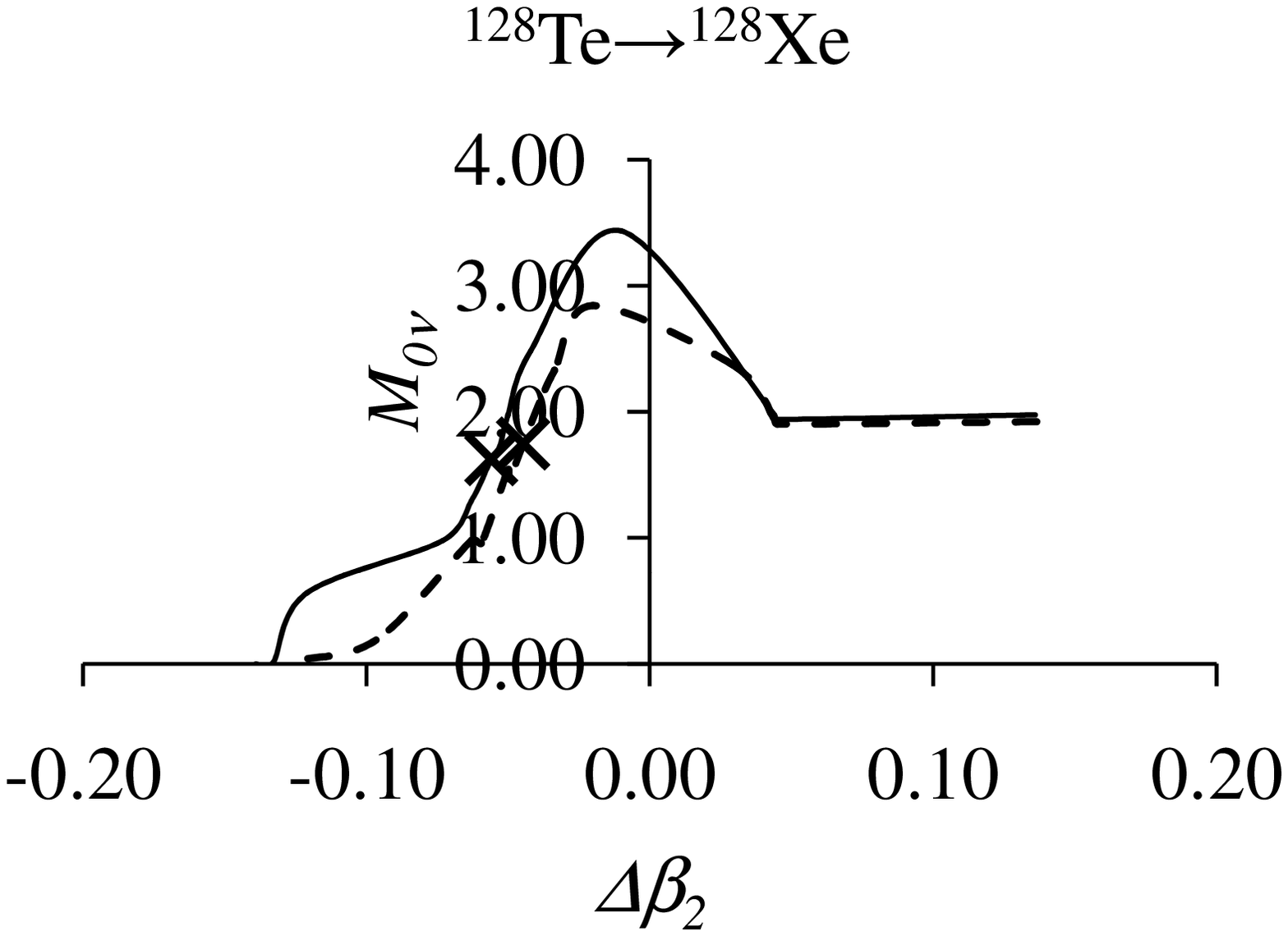}} \\
\resizebox{0.215\textwidth}{!}{\includegraphics{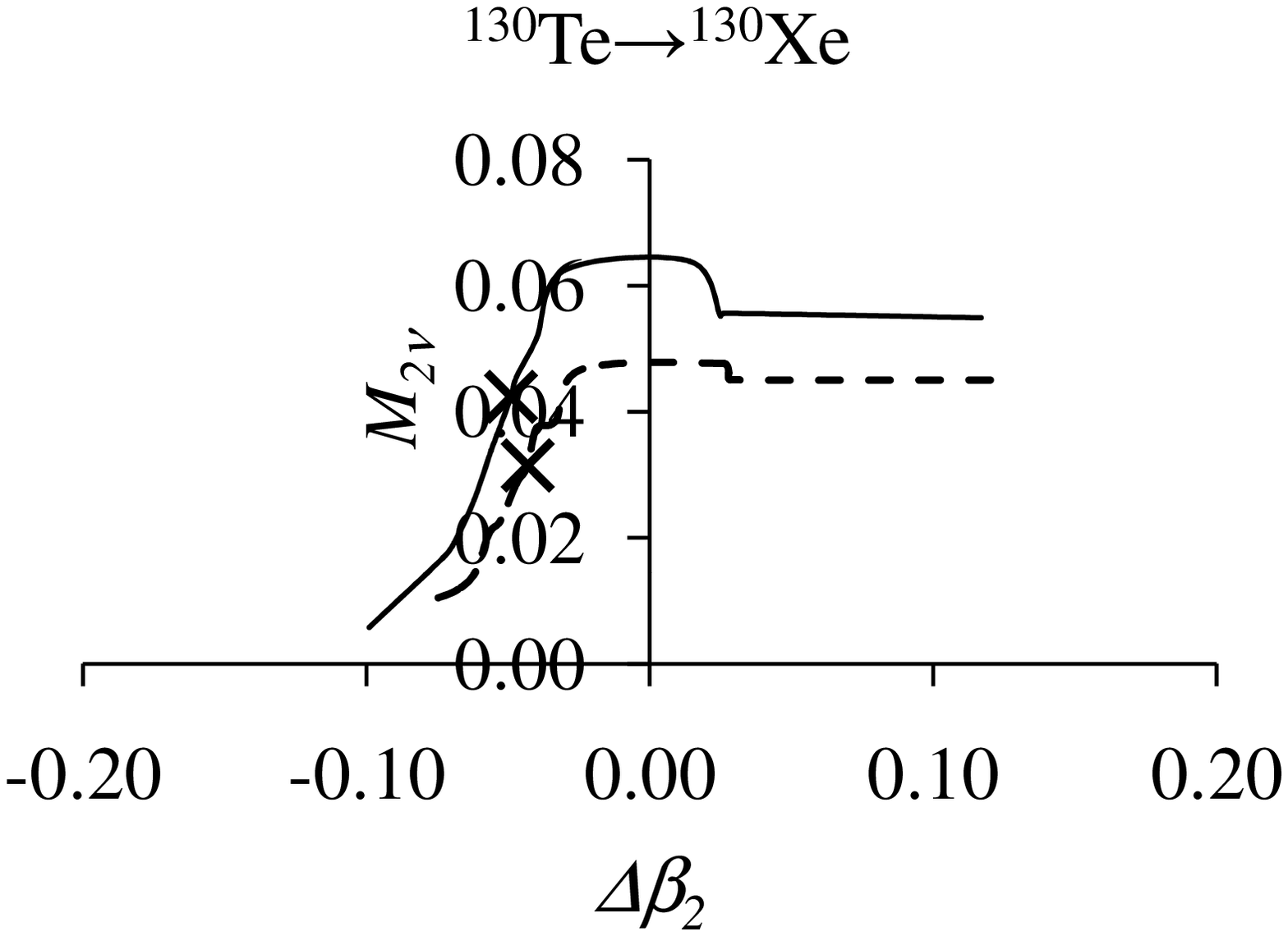}} &
\resizebox{0.215\textwidth}{!}{\includegraphics{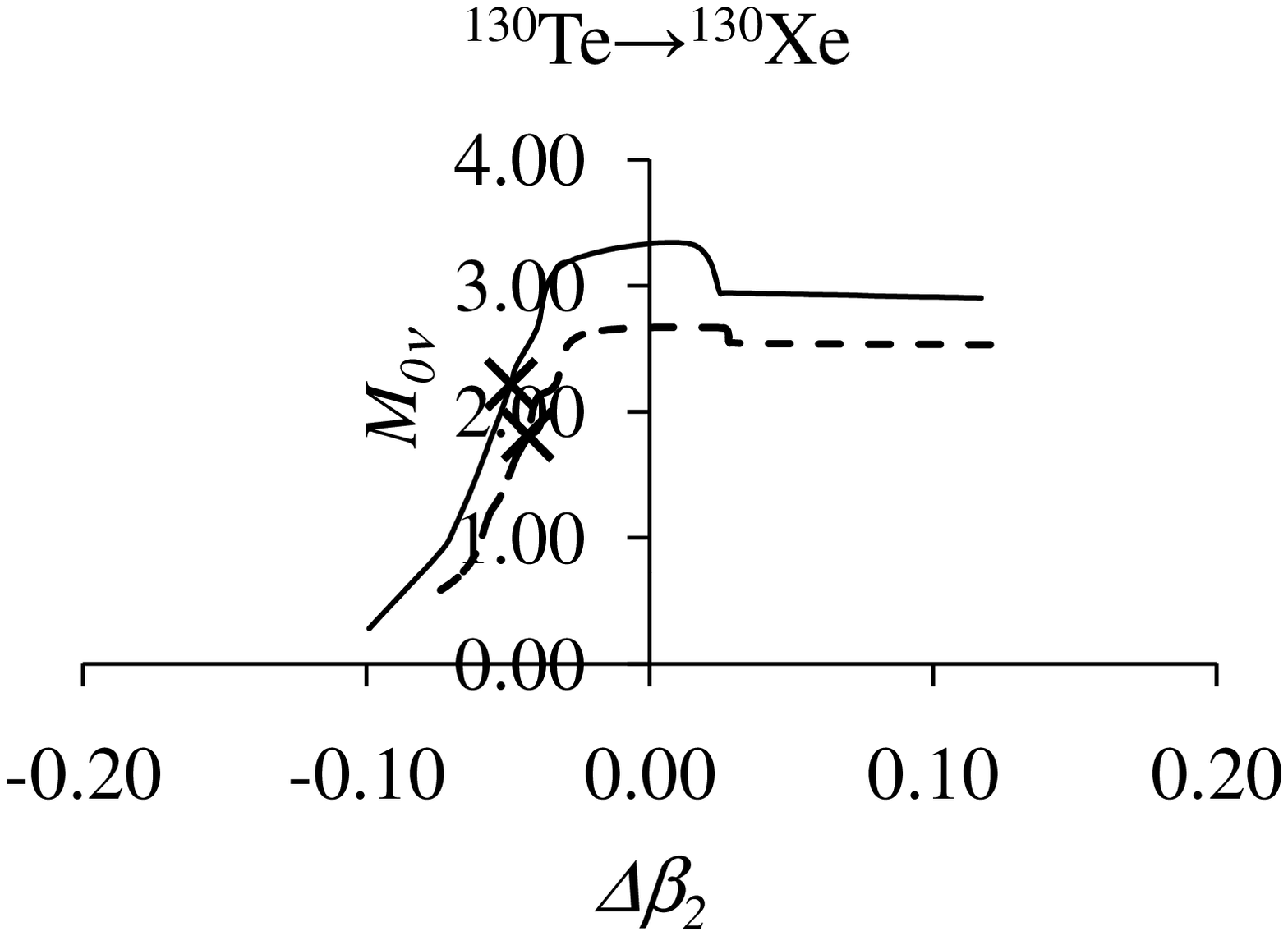}} \\
\resizebox{0.215\textwidth}{!}{\includegraphics{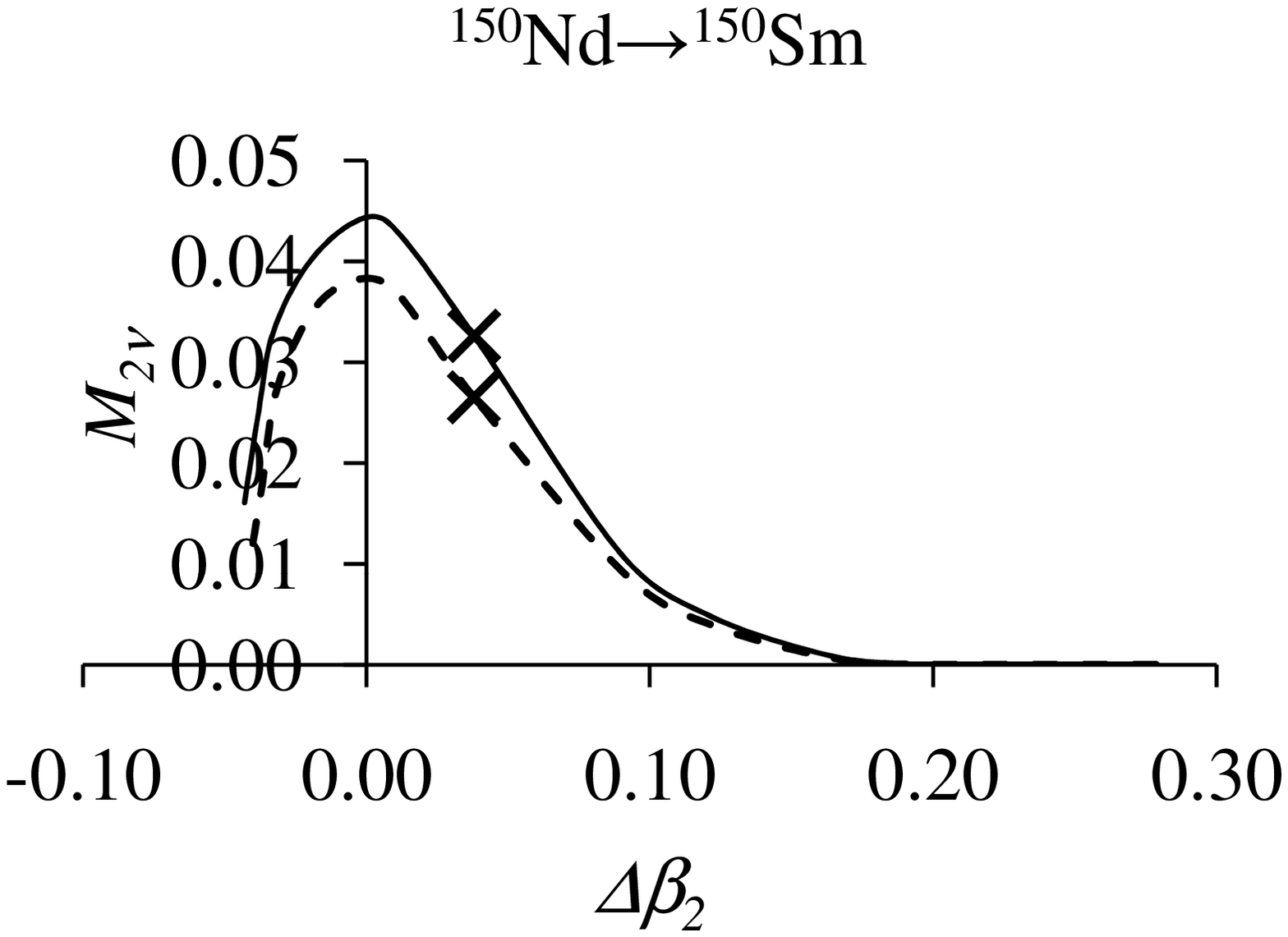}} &
\resizebox{0.215\textwidth}{!}{\includegraphics{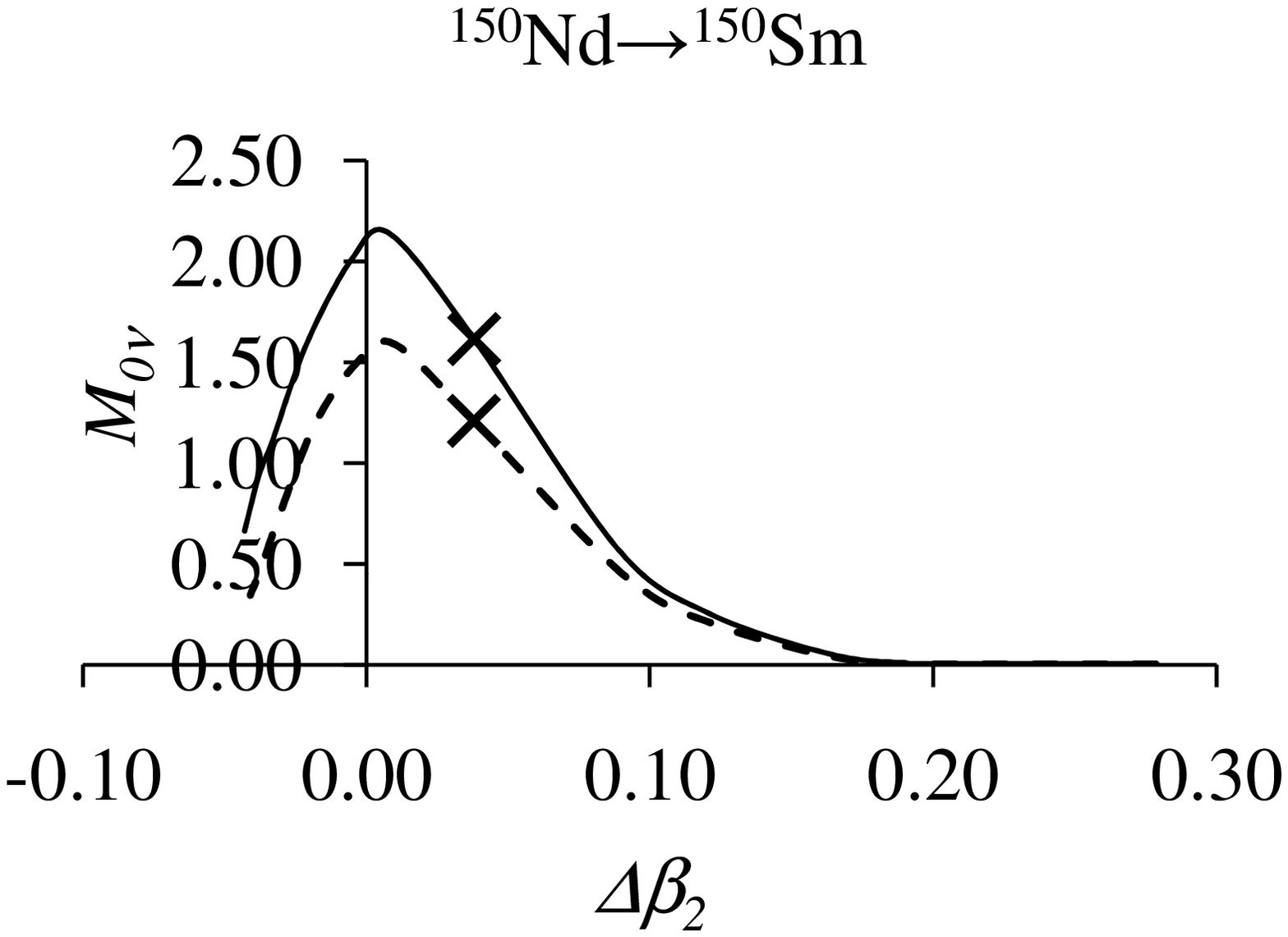}}
\end{tabular}
\caption{NTMEs of $^{128,130}$Te and $^{150}$Nd nuclei
for $\left(\beta^{-}\beta^{-}\right)_{2\nu}$ 
and
$\left(\beta^{-}\beta^{-}\right)_{0\nu}$ decay. Further details are 
given in fig. 1.}
\label{fig2}
\end{figure}

A suppression of NTMEs $M_{2\nu }$ for $\beta \beta $ decay with respect
to the spherical case has been reported when the parent and daughter nuclei
have different deformations \cite{alva04,mene08}. To investigate this
effect, we present the NTMEs for the $\left( \beta ^{-}\beta ^{-}\right)
_{2\nu }$ as well as $\left( \beta ^{-}\beta ^{-}\right) _{0\nu }$ decay of $%
^{94,96}$Zr, $^{98,100}$Mo, $^{104}$Ru, $^{110}$Pd isotopes in fig.~\ref{fig1}
and $^{128,130}$Te and $^{150}$Nd isotopes in fig.~\ref{fig2} as a function 
of the difference in the deformation parameter $\Delta \beta _{2}=\beta
_{2}(parent)-\beta _{2}(daughter)$ between the parent and daughter nuclei.
The NTMEs are calculated by keeping the deformation for parent nuclei fixed
at $\zeta _{qq}=1$ and 
changing the deformation of daughter nuclei by varying $\zeta _{qq}$ in the range 0.0 -- 1.5.\footnote{
%
In table 5 of ref. \cite{sing07}, the given $\beta _{2}$ values of $^{130}$%
Xe (for $^{130}$Te$\rightarrow ^{130}$Xe) at $\zeta _{qq}=0.9$ and 1.5 are
0.159 and 2.155 respectively, which is a typographical error. 
The correct values are 0.157 and 0.216 respectively.} 

In our earlier works \cite{chan05,sing07}, it has been shown that the NTMEs $%
M_{2\nu }$ are usually large for $\zeta _{qq}=0.0$ i.e. when both the parent
and daughter nuclei are spherical. With the increase of $\zeta _{qq}$, the
NTMEs remain almost constant and then decrease around the physical value $%
\zeta _{qq}=1.0$ establishing an inverse correlation between $M_{2\nu }$ and 
$\beta _{2}$. 
It can be observed at the left panels of figs.~\ref{fig1} and \ref{fig2} that in all cases 
that when the absolute value of the difference in deformation is close to zero
the largest NTMEs are obtained. In addition, the effect of hexadecapolar correlations 
is almost negligible.
%

In the mass mechanism neglecting pseudo-scalar and weak magnetism terms
of the recoil current, the importance of which has been reported by
Simkovic \etal \cite{simk99,verg02} and needs further investigation, 
the inverse half-life of 
the $\left( \beta ^{-}\beta ^{-}\right) _{0\nu }$
 decay due to the exchange of light neutrinos for the$%
\ 0^{+}\to 0^{+}$\ transition is given by \cite{haxt84,doi85,tomo91} 
\begin{equation}
\left[T_{1/2}^{0\nu }(0^{+}\to 0^{+})\right]^{-1}=\left(\frac{\left\langle
m_{\nu }\right\rangle }{m_{e}}\right)^{2}G_{01}|M_{0\nu }|^{2}
\end{equation}
where the NTME $M_{0\nu }$ is given by
{\footnotesize
\begin{equation}
M_{0\nu }=\sum_{n,m}\left\langle 0_{F}^{+}\left\| \left[ \mathbf{\sigma }%
_{1}\cdot \mathbf{\sigma }_{2}-\left( \frac{g_{V}}{g_{A}}\right) ^{2}\right]
H_{m}(r)\tau _{n}^{+}\tau _{m}^{+}\right\| 0_{I}^{+}\right\rangle
\end{equation}
}
The calculation of $M_{0\nu }$ in the PHFB model has been discussed in 
an
earlier work \cite{chat08}. It is worth mentioning that we include the finite
size and short range effects through a dipole form factor and Jastrow type 
of short range correlation (SRC). The other two prescriptions for including SRC namely,
through the exchange of $\rho$ and $\omega$ mesons \cite{hirs95} and
unitary correlation operator method \cite{kort07,simk08} need to be
investigated.
In table~\ref{tab2}, we give the theoretically calculated 
$M_{0\nu }$ using the HFB wave functions in conjunction with 
$PQQ$ \cite{chat08} and
$PQQHH$ interaction. The change in the values of NTMEs $M_{0\nu }$ are below 10\%
except $^{110}$Pd, $^{130}$Te and $^{150}$Nd, for which the changes are 17$%
\% $, 18$\%$ and 25$\%$ respectively. The ratio for deformation effect $%
D_{0\nu }$ are 2.51(2.30), 4.50(4.21), 1.94(1.87), 2.15(1.92), 3.81(3.85),
2.64(2.87), 4.43(3.85), 2.96(3.26) and 6.16(7.14) for $^{94,96}$Zr, $%
^{98,100}$Mo, $^{104}$Ru, $^{110}$Pd, $^{128,130}$Te and $^{150}$Nd nuclei
with $PQQ$($PQQHH$) interaction respectively corresponding to a maximum
change up to 16\%.
It is worth mentioning that, 
the $\beta_2$ for $^{150}$Nd is close to the experimental value whereas 
for $^{150}$Sm, it is off. For the $PQQ$ interaction, the experimental $\beta_2$ but for other physical properties can be reproduced with 
$\zeta_{qq}=0.96$ and the corresponding $M_{0\nu}$ is 0.7015.

In the right panel of figs.~\ref{fig1} and \ref{fig2}, we present the 
variation of $\left|M_{0\nu }\right| $ with respect to 
$\left| \Delta \beta _{2}\right| $ for
the above mentioned $\beta ^{-}\beta ^{-}$ emitters. It is noticed that
variation in $\left| M_{0\nu }\right| $ with changing $\left| \Delta \beta
_{2}\right| $ is similar to that of $\left( \beta ^{-}\beta ^{-}\right)
_{2\nu }$ decay. Moreover, it is noticed from figs.~\ref{fig1} and \ref{fig2} 
that the NTMEs remain constant even when one of the nuclei is spherical 
or slightly deformed. With further increase in deformation, the NTMEs 
become the maximum for $\left| \Delta \beta _{2}\right| =0$ and then 
decrease with increase in the difference between the deformation parameter. 

We conclude that the effect of including the hexadecapole interaction channel is minor 
both on the deformation $\beta_2$ and on the NTMEs. In particular, it is clear that the 
qualitative dependence of the NTMEs on deformation is not affected by including
the $HH$ interactions in the Hamiltonian, and the changes, while visible, are small. It exhibits that the 
inclusion of $QQ$ interaction is a necessary ingredient for a reliable 
calculation of NTMEs, and that effect of the hexadecapolar correlations can be 
mimicked by a slight renormalization in the $QQ$ channel. 
%
The independent deformations of initial and final nuclei are,
indeed, crucial 
parameters to describe realistic NTMEs
$M_{2\nu }$ and $M_{0\nu }$ of $\beta \beta $ decay,
which exhibit, in their respective scales, a very similar dependence on the difference of deformations.

\acknowledgments
Partial grant by DST, India (SR/S2/HEP-13/2006), Conacyt-M\'{e}xico and DGAPA-UNAM is acknowledged.

\end{document}